\documentclass{aa}
\usepackage{latexsym}
\usepackage{epsfig}
\newcommand{\radm}{rad~m$^{-2}$} 

\newcommand{\dg}{\degr}

\begin{document}

\title{Characteristics of the structure in the Galactic polarized
	  radio background at 350 MHz}

\author{M. Haverkorn\inst{1}
        \and
	P. Katgert\inst{2}
	\and
	A. G. de Bruyn\inst{3,4}
        }
\offprints{M. Haverkorn}
\institute{Leiden Observatory, P.O.Box 9513, 2300 RA Leiden, the
              Netherlands\\
	      (Current address: Harvard-Smithsonian Center for
	      Astrophysics, 60 Garden Street MS-67, Cambridge MA
	      02138, USA)\\
              \email{mhaverkorn@cfa.harvard.edu} 
         \and Leiden Observatory, P.O.Box 9513, 2300 RA Leiden, the
              Netherlands\\ 
              \email{katgert@strw.leidenuniv.nl} 
         \and ASTRON, P.O.Box 2, 7990 AA Dwingeloo, the Netherlands\\ 
              \email{ger@astron.nl}
         \and Kapteyn Institute, P.O.Box 800, 9700 AV Groningen,
              the Netherlands }
\titlerunning{Structure in the Galactic polarized radio background}

\abstract{
Angular power spectra and structure functions of the Stokes parameters
$Q$ and $U$ and polarized intensity $P$ are derived from three sets
of radio polarimetric observations. Two of the observed fields have
been studied at multiple frequencies, allowing determination of power
spectra and structure functions of rotation measure $RM$ as well. The
third field extends over a large part of the northern sky, so that the
variation of the power spectra over Galactic latitude and longitude
can be studied.  The power spectra of $Q$ and $U$ are steeper than
those of $P$, probably because a foreground Faraday screen creates
extra structure in $Q$ and $U$, but not in $P$. The extra structure in
$Q$ and $U$ occurs on large scales, and therefore causes a steeper
spectrum. The derived slope of the power spectrum of $P$ is the
multipole spectral index $\alpha_P$, and is consistent with earlier
estimates. The multipole spectral index $\alpha_P$ decreases with
Galactic latitude (i.e.\ the spectrum becomes flatter), but is
consistent with a constant value over Galactic longitude. Power
spectra of the rotation measure $RM$ show a spectral index
$\alpha_{RM} \approx 1$, while the structure function of $RM$ is
approximately flat. The structure function is flatter than earlier
estimates from polarized extragalactic sources, which could be due to
the fact that extragalactic source $RM$ probes the complete line of
sight through the Galaxy, whereas as a result of depolarization
diffuse radio polarization only probes the nearby ISM.
}

\maketitle

\section{Introduction}

The warm ionized gaseous component of the Galactic interstellar medium
(ISM) shows structure in density and velocity on scales from AU to
several kiloparsecs.  The Galactic magnetic field is coupled to the
motions of the ionized gas and has a comparable energy density, so
that gas and magnetic field are in complex interaction.  Detailed
knowledge of the turbulent nature of the warm ISM and the structure in
the Galactic magnetic field is essential for several fundamental
studies of the Galaxy, including modeling of molecular clouds
(e.g. V\'azquez-Semadeni \& Passot 1999, Ostriker et al.\ 2001),
heating of the ISM (Minter \& Balser 1997), star formation
(e.g. Ferri\`ere 2001), and cosmic ray propagation (Chevalier \&
Fransson 1984). 

Small-scale structure in the warm ISM and magnetic field can be well
studied using radio polarimetric observations, of the synchrotron
background in the Milky Way (e.g.\ Brouw \& Spoelstra 1976, Wieringa
et al.\ 1993, Duncan et al.\ 1997, 1999, Uyan\i ker et al.\ 1999,
Landecker et al.\ 2001, Gaensler et al.\ 2001), of pulsars (e.g.\ Rand
\& Kulkarni 1989, Ohno \& Shibata 1993, Rand \& Lyne 1994, Han et
al.\ 1999) or of polarized extragalactic point sources (e.g.\
Simard-Normandin \& Kronberg 1980, Clegg et al.\ 1992). At short
wavelengths ($\lambda \la 6$~cm), Faraday rotation is negligible, so
that the measured polarization directly traces the magnetic field in
the emitting region. At longer wavelengths, Faraday rotation
measurements give additional information on density and magnetic field
structure along the entire line of sight. Furthermore, depolarization
processes define a distance beyond which polarized radiation is
significantly depolarized, which depends on wavelength. So
high-frequency measurements probe the total line of sight through the
Galaxy, whereas low-frequency polarization observations only trace the
nearby part of the ISM.

Specific intriguing small-scale structures and discrete objects have
been studied in diffuse polarization observations (Gray et al.\ 1998,
Haverkorn et al.\ 2000, Uyan\i ker \& Landecker 2002), but the
spatial structures have also been analyzed statistically. Simonetti et
al.\ (1984) and Simonetti \& Cordes (1986) studied the structure
in Galactic rotation measure $RM$ by comparing $RM$s of polarized
extragalactic sources, and of separate components of the same
source. The electron density seems to exhibit a power law structure
function and therefore a power law angular spectrum (see 
also Armstrong et al.\ 1995, Minter \& Spangler 1996). Recently,
statistical analysis of the diffuse Galactic polarized foreground have
been pursued in the form of angular power spectrum studies (Tucci et
al.\ 2000, 2002, Baccigalupi et al.\ 2001, Giardino et al.\ 2002,
Bruscoli et al.\ 2002), with the objective of estimating the
importance of the Galactic ISM as a foreground contaminator for CMBR
polarization observations (e.g. Seljak 1997, Prunet et al.\ 2000).

In this paper, we study the statistical properties of the warm ISM and
Galactic magnetic field by means of power spectra of Stokes parameters
$Q$ and $U$, polarized intensity $P$ and rotation measure $RM$. We
also derive the structure function of $RM$ to allow a comparison with
earlier studies of $RM$ structure functions from polarized
extragalactic sources. Furthermore, by careful selection of reliable
$RM$ determinations (i.e.\ those with low $\chi^2$ of the linear fit
to $\phi(\lambda^2)$) in the calculation of structure functions, we
can obtain an estimate of how much the structure functions (and power
spectra) are influenced by low-quality $RM$s.

We use data from three regions, all at positive Galactic latitudes, in
which we observed the diffuse polarized emission at frequencies around
350~MHz with the Westerbork Synthesis Radio Telescope (WSRT). The
first two regions, each about 50 square degrees in size, in the
constellations Auriga (Haverkorn et al.\ 2003a) and Horologium
(Haverkorn et al.\ 2003b), are observed at multiple frequencies at a
resolution of $\sim$~5\arcmin. The third region is a part of the
Westerbork Northern Sky Survey (WENSS, Rengelink et al.\ 1997), a
high-resolution radio survey at 327~MHz of the northern
hemisphere (Schnitzeler et al., in prep). For those parts of the WENSS
survey that were observed at night, polarization data are usable. Here
we discuss polarization data from the region with $140\dg
\la l \la 170\dg$ and $0\dg \la b \la 30\dg$.

In Sect.~\ref{s9:data} we describe the three sets of data that we
analyze in this paper. In Sect.~\ref{s9:ps} angular power spectrum
analysis is introduced, power spectra of the data are presented and
discussed and literature on the angular power spectra is briefly
summarized. Section~\ref{s9:sf} gives structure functions for $RM$ in
the two multi-frequency measurements. In Sect.~\ref{s9:disc} the
results are discussed, and finally in Sect.~\ref{s9:conc} some
conclusions are stated. 

\section{The observations}
\label{s9:data}

\subsection{Multi-frequency WSRT observations}
\label{s9:wsrtdata}

\begin{figure*}
  \centering
  \hbox{\psfig{figure=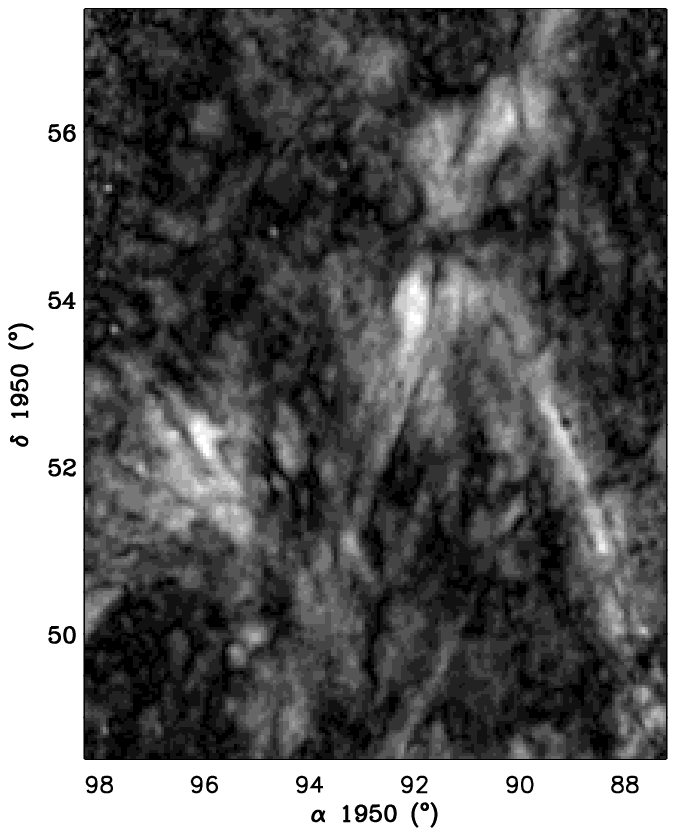,width=.5\textwidth}
        \psfig{figure=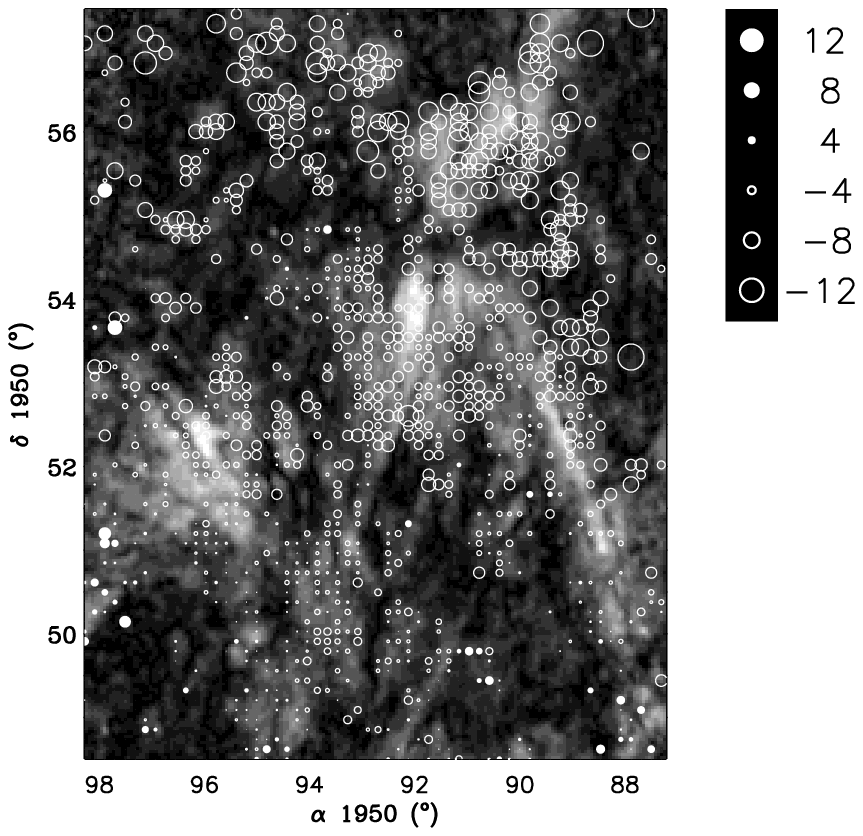,width=.5\textwidth}}
  \caption{Grey scale representation of the polarized intensity $P$ at
  349~MHz and rotation measure $RM$, of the region in the
  constellation of Auriga centered at $(l,b) = (161\dg,16\dg)$ at a
  resolution of $\sim 5\arcmin$. The grey scale is saturated at $P =
  95$~mJy/beam (white) ($T_{b,pol} = 12$~K). The right panel shows
  $RM$ in superimposed circles, where the diameters of the
  circles indicate $RM$, and filled circles denote positive $RM$s. The
  scaling of $RM$ on the right-hand side is in \radm. }
  \label{f9:pirm_aur}
\end{figure*}

We carried out low frequency radio polarimetry with the Westerbork
Synthesis Radio Telescope (WSRT), in two regions of the sky at
positive latitudes, in the constellations of Auriga and
Horologium. Data were obtained simultaneously at 8 frequencies around
350~MHz, each with a bandwidth of 5~MHz. Due to radio interference and
hardware problems, only data in 5 frequency bands could be used, viz.\
those centered at the frequencies 341, 349, 355, 360, and 375~MHz. The
multi-frequency data allow the study of the frequency dependence of
the polarization structure, and the determination of rotation measure
$RM$. We use the technique of mosaicking (i.e.\ the telescopes cycle
through a number of adjacent fields on the sky during a 12hr
observation) to obtain a field of view that is larger than the primary
beam. Mosaicking also suppresses instrumental polarization to below
1\% (see Haverkorn 2002). Maps of the Stokes parameters $I$, $Q$, and
$U$ were derived from the observed visibilities.  $RM$s were computed
straightforwardly from the linear relation between polarization angle
$\phi$ and $\lambda^2$. Because the observed $RM$ values are small
($|RM| \la 10$~\radm), there is no $n\,180$\dg\ ambiguity in $\phi$,
and $RM$s can be computed with $|\phi(\lambda_i) - \phi(\lambda_j)| <
90\dg$ for adjacent wavelengths $\lambda_i$ and $\lambda_j$.  We
define a determination of $RM$ in a particular beam reliable if (1)
the reduced $\chi^2$ of the linear $\phi(\lambda^2)$-relation
$\chi^2_{red} < 2$, and (2) the polarized intensity averaged over
wavelength $\left<P\right> > 20$~mJy/beam (i.e. $\sim 4\sigma$).

The maximum baseline of the observations was 2700m, yielding a
resolution of 1\arcmin, but smoothing of the Stokes $Q$ and $U$ data
(using a Gaussian taper in the $(u,v)$-plane) was applied to obtain a
better signal-to-noise.  The taper has a value of 0.25 at a baseline
value around 300m, where the exact values were chosen so that the beam
size is identical at all 5 frequencies, viz.\
5.0\arcmin$\times$5.0\arcmin~cosec~$\delta$.

The first region, in the constellation Auriga, is centered on $(l,b) =
(161\dg, 16\dg)$ and is about $9\dg\times 7\dg$ in size.  The left
panel in Fig.~\ref{f9:pirm_aur} shows the polarized intensity $P$ at
349~MHz in the Auriga region at 5.0\arcmin\ resolution. We do not show
the original 1\arcmin\ resolution map, because it is
noise-dominated. The maximum polarized brightness temperature in the
map is $T_{b,pol} \approx$~13~K, and the noise is about 0.45~K.
Rotation measures in the Auriga region are shown in the right panel of
Fig.~\ref{f9:pirm_aur} as circles superimposed on the grey scale map
of $P$ at 349~MHz. The diameter of each circle indicates the value of
the $RM$ at that position, where filled circles denote positive $RM$s,
and open circles negative $RM$s.  We show only the $RM$s that are
reliably determined, and only one in four independent beams.

\begin{figure*}
  \begin{center}
    \hbox{\psfig{figure=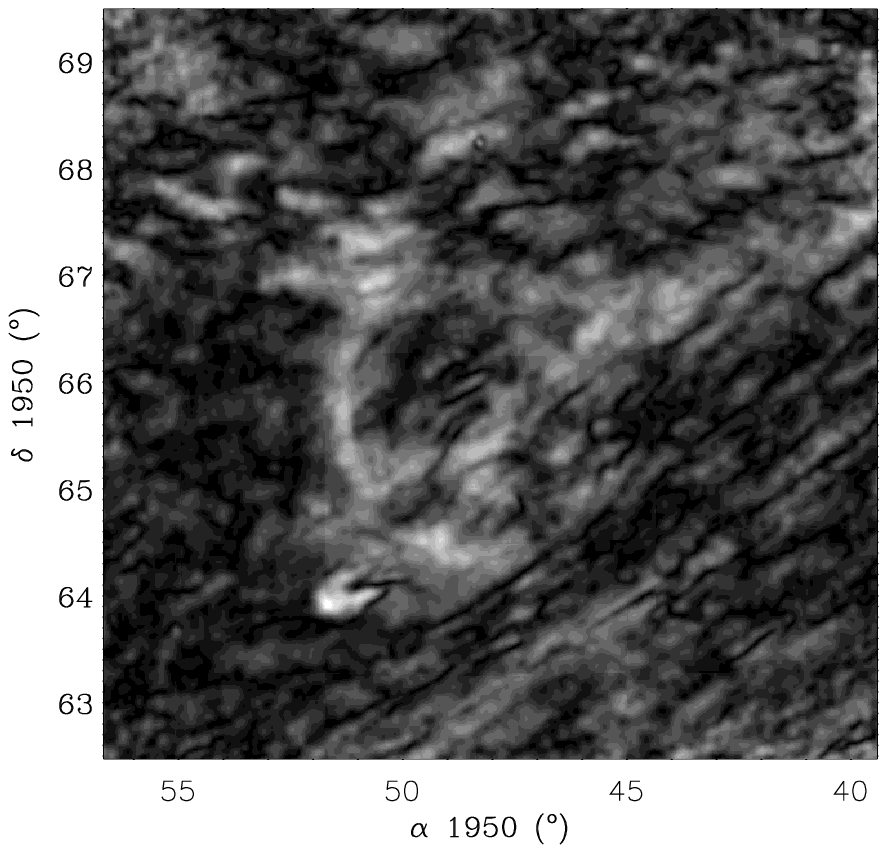,width=.5\textwidth}
          \psfig{figure=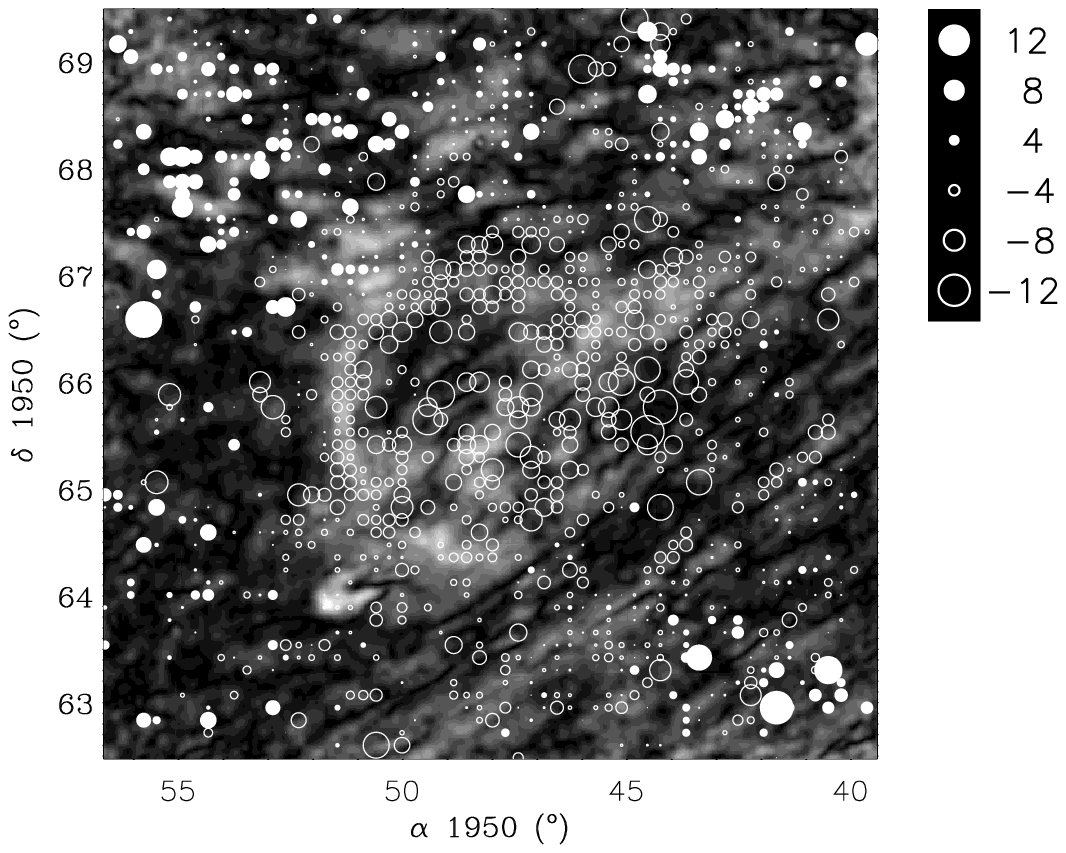,width=.5\textwidth}}
    \caption{$P$ and $RM$ for the region in the constellation of
      Horologium centered at $(l,b) = (137\dg,7\dg)$. Notation as in
      Fig.~\ref{f9:pirm_aur}, where the maximum $P = 95$~mJy/beam
      corresponds to $T_{b,pol} = 13.9$~K.}
  \label{f9:pirm_hor}
  \end{center}
\end{figure*}

The region in the constellation Horologium, centered at $(l,b) =
(137\dg,7\dg)$, was observed in the same way as the Auriga region, at
the same five frequencies, and with the same taper applied.
Fig.~\ref{f9:pirm_hor} shows the polarized intensity $P$ at 349~MHz in
the left panel, and $RM$s in the form of overlaid circles in the right
panel. $RM$s are coded in the same way as in Fig.~\ref{f9:pirm_aur}.
The maximum polarized brightness temperature is $T_{b,pol}
\approx$~17~K, and the noise is a little higher than that in the
Auriga field, about 0.65~K.

The polarized intensity is not corrected for noise bias
$P_{debias} = \sqrt{(Q_{obs}^2 + U_{obs}^2) - \sigma^2}$ (for $P >
\sigma$). However, $P$ generally has a S/N~$> 4-5$, for which the
debiasing does not alter the data by more than 2~-~3\%. Furthermore,
power spectra are not affected by debiasing except at zero frequency.

In both fields there is no small-scale structure visible in the map of
total intensity Stokes $I$, despite the ubiquitous structure on
arcminute and degree scales in $P$. Because the large-scale ($\ga
1\dg$) component of $I$ cannot be measured with the WSRT due to
missing short spacings, we used the Haslam et al.\ (1981, 1982) radio
survey at 408~MHz to estimate the total intensity $I$ to be
$\sim$~34~K and $\sim$~47~K in Auriga and Horologium, respectively. The
polarized intensity $P$ shows structure up to $\sim$~10 - 15~K on
scales from arcminutes to degrees. Due to the lack of corresponding
structure in $I$, the structure in polarized intensity cannot be
due to small-scale synchrotron emission but must be due to
other, instrumental and depolarization, mechanisms. 

In a medium that emits synchrotron radiation and simultaneously causes
Faraday rotation, the polarized emission is depolarized by so-called
depth depolarization, due to the vector averaging of contributions
from different parts of the line of sight. This, together with beam
depolarization (due to angle structure within one synthesized beam),
produces structure in $P$. Furthermore, the insensitivity of the
interferometer to large-scale structure can cause additional structure
in $P$. However, we have shown that this effect cannot be very
important in these observations (Haverkorn et al.\ 2003c).

\subsection{Polarization data from the WENSS survey}

The Westerbork Northern Sky Survey (WENSS, Rengelink et al.\ 1997) is
a low-frequency radio survey that covers the whole sky north of
$\delta=30$\dg\ at  325~MHz to a limiting flux density of
approximately 18~mJy (5$\sigma$) and with a resolution of
54\arcmin\arcmin$\times$54\arcmin\arcmin~cosec~$\delta$.

Polarization data taken during the day are greatly affected by solar
radiation, which is detected in sidelobes. Furthermore, ionospheric
Faraday rotation rapidly changes during sunrise and sunset, causing a
reduction of the apparent polarized intensity, and adding much noise.
However, in mosaics observed (almost) entirely during night time, the
polarization data are of good quality. As a result, we could make a
``supermosaic'' of a large region of $\sim 30\dg\times35\dg$, which we
will refer to as the WENSS polarization region. Details on the
observations and correction for ionospheric Faraday rotation are given
in Schnitzeler et al.\ (in prep).

\begin{figure*}
  \centering
  \psfig{figure=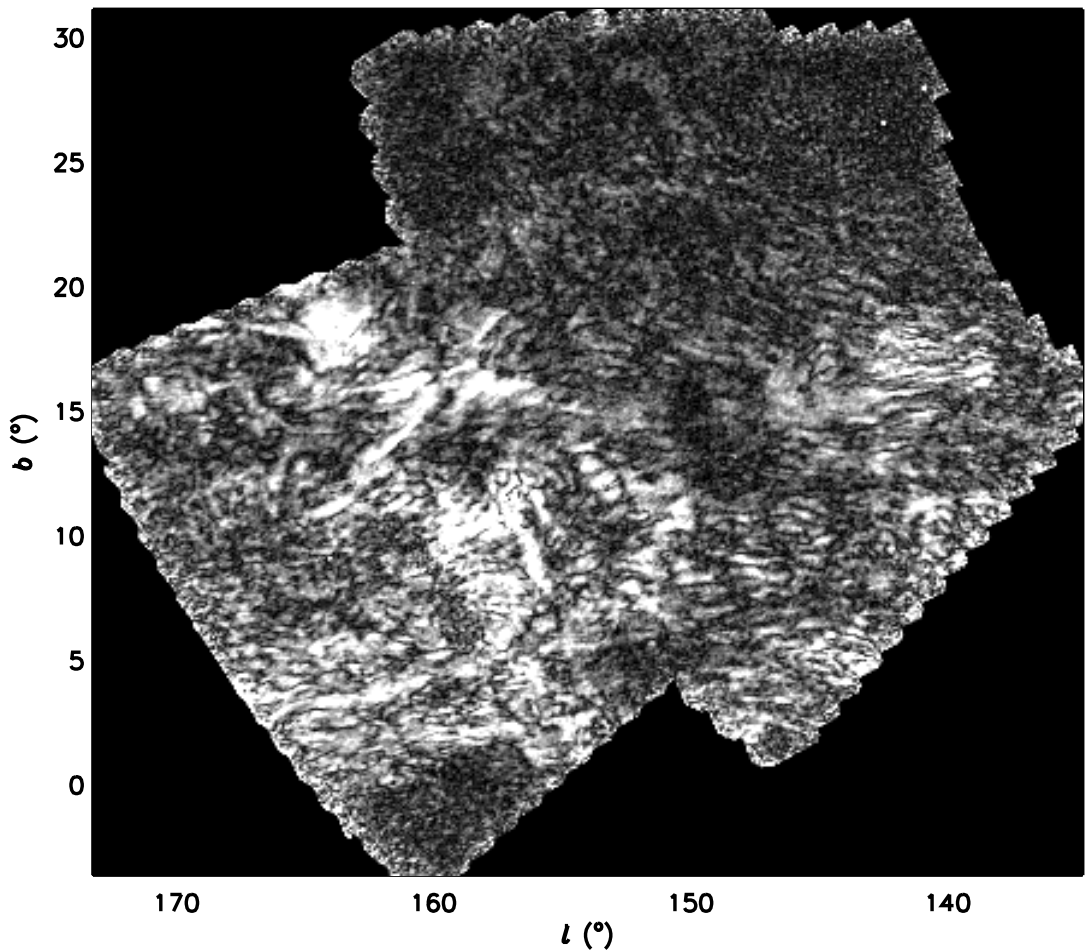,width=.8\textwidth}
  \caption{Grey scale representation of the polarized intensity of
      the polarization part of the WENSS survey at 325~MHz, where
      white denotes a maximum intensity of 35~mJy/beam ($T_{b,pol}
      \approx$ 17~K to 25~K, depending on declination). The data were
      smoothed with a 500m taper, and have a resolution of
      $\sim$~2.5\arcmin.}
  \label{f9:wenss}
\end{figure*}

Fig.~\ref{f9:wenss} shows polarized intensity $P$ in the WENSS
polarization region, resampled in Galactic longitude and latitude. For
this analysis, a Gaussian taper with value 0.25 at a baseline of 500m
was applied, which yields a resolution of $\sim$~2.5\arcmin. In the
figure, $P$ saturates at 35~mJy/beam, which coincides with a polarized
brightness temperature of $\sim$~17~K to $\sim$~25~K, depending on
declination. The average $P \approx 2.6$~mJy/beam. Note that these
observations are taken in a single 5~MHz wide frequency band, so
rotation measure data are not available.

\section{Angular power spectrum analysis}
\label{s9:ps}
 
\subsection{Determination of the multipole spectral index}

To quantify the structure in the polarization maps, we calculate
angular power spectra $PS (\ell)$ as a function of multipole $\ell$. A
multipole $\ell$ is a measure of angular scales equivalent to wave
number, and is defined as $\ell \approx 180\dg /\theta$, where
$\theta$ is the angular scale in degrees. The angular power spectrum
$PS$ of a radiation field $X$ is the square of the Fourier transform
of $X$: $PS_X (\ell) = |{\cal F}(X)|^2$, where ${\cal F}$ denotes the
Fourier transform.  Here, the observable $X$ can be either Stokes $Q$,
Stokes $U$, polarized intensity $P$ or rotation measure $RM$.  The
power spectra were computed in two dimensions, and averaged over
azimuth in radial bins.  The multipole spectral index $\alpha$,
defined as $PS_X (\ell) \propto \ell^{-\alpha}$, is calculated from a
log-log fit to the power spectrum. In the tapered data, multipoles
with higher values of $\ell$ are affected by the tapering, and in the
untapered data higher multipoles are dominated by noise.  Multipoles
with $\ell \la 200$ correspond to angular scales $\theta \ga
1$\dg, to which the WSRT is not sensitive.

The visibilities from which a map is made are $V_{map}(u,v) =
V_{obs}(u,v)\, T(u,v)$, where $V_{obs}$ are the observed visibilities,
and $T(u,v)$ is the taper function. The calculated intensity of
the tapered data is ${\cal F}(V_{map}) = {\cal F}(V_{obs})*{\cal F}(T)$,
where ${\cal F}$ is a Fourier transform and the asterisk denotes
convolution. The power spectrum of the Stokes parameter $X$, $PS(X)$,
is then:
\begin{equation}
  PS_X(\ell) = |{\cal F}(X)|^2 = |{\cal F}(X_{obs})|^2 \, T^2 =
             PS_{X,obs}(\ell) \, T^2
  \label{e:taper}
\end{equation}
where $X$ is Stokes $Q$ or $U$. Although polarized intensity $P$
is derived from $Q$ and $U$ and thus not directly observed, correction
for the taper in the same way as for $Q$ and $U$ power spectra is a
good approximation. As an illustration, Fig.~\ref{f9:ps_corr} shows
the power spectrum of $P$ of the tapered data in the Auriga region at
341~MHz (solid line). The dotted line is the same spectrum, but
corrected for the tapering according to Eq.~(\ref{e:taper}). The power
law behavior extends to $\log(\ell) \approx 3.6$. 

\subsection{Power spectra from the multi-frequency WSRT studies}

\begin{figure}
  \centering
  \psfig{figure=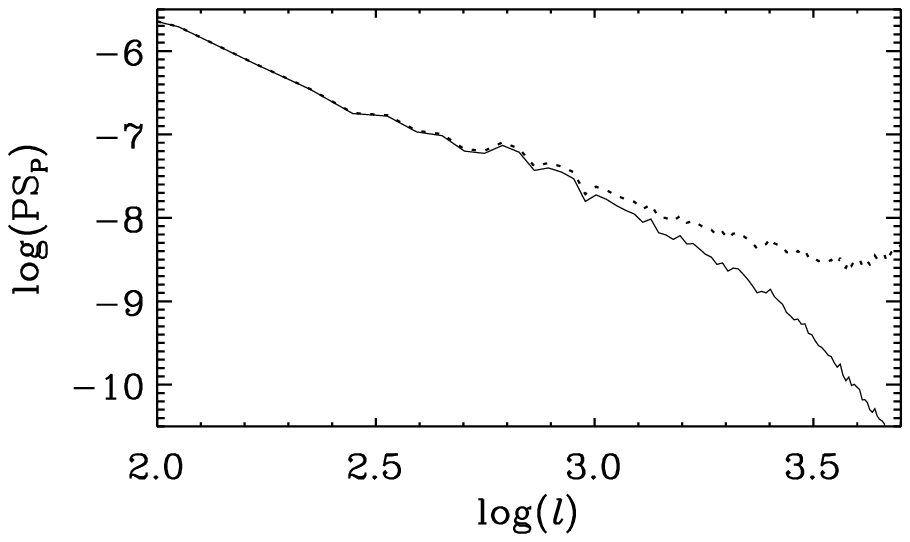,width=.4\textwidth}
  \caption{Power spectrum of $P$ at 341~MHz in the Auriga field. The
      solid line shows the observed data, the dotted line is corrected
      for the 300m taper.\vspace*{2.65cm}}
  \label{f9:ps_corr}
\end{figure}

\begin{figure}
  \centering
  \psfig{figure=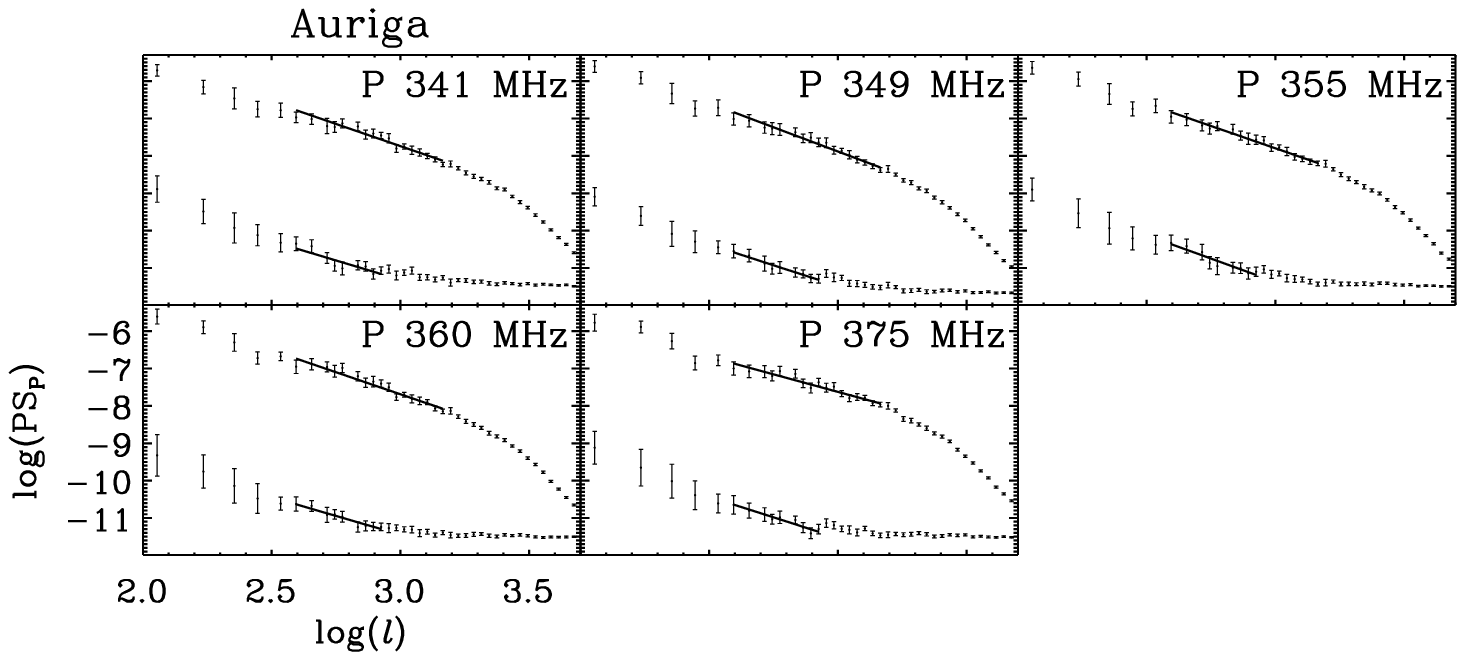,width=.5\textwidth}
  \psfig{figure=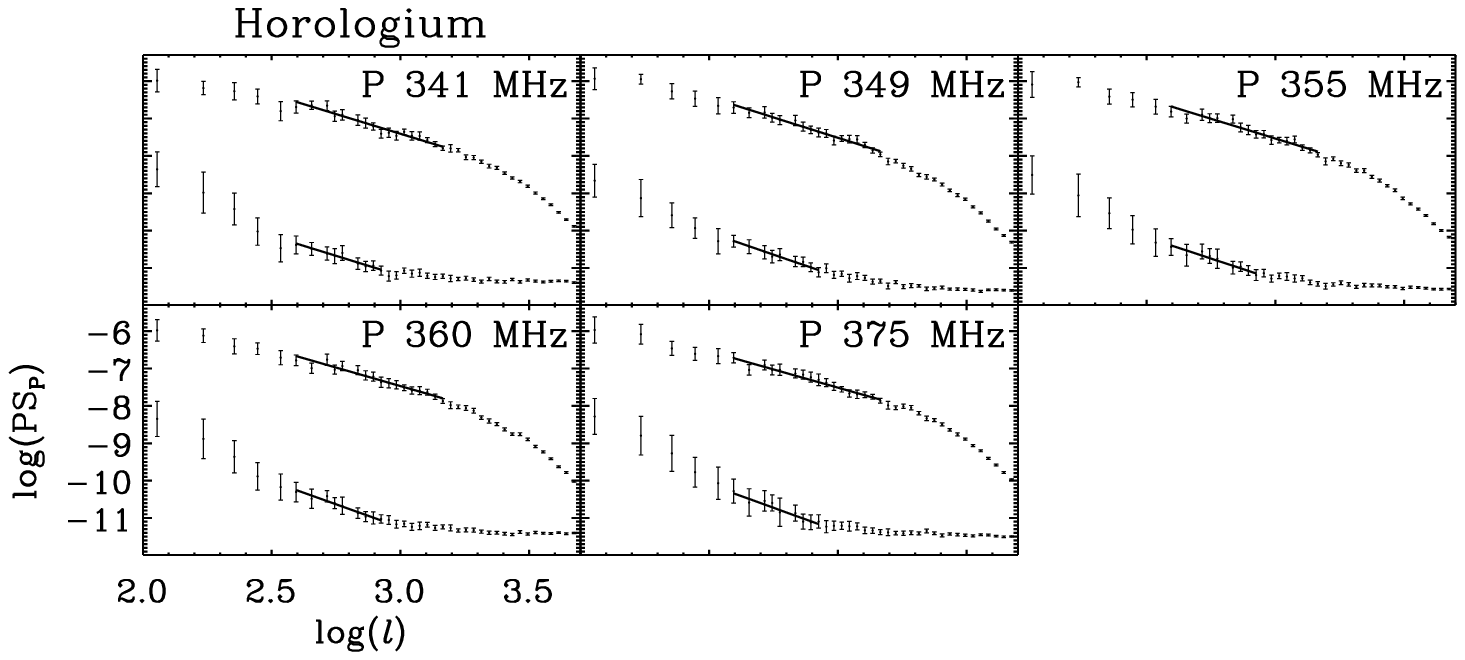,width=.5\textwidth}
  \caption{Power spectra of polarized intensity $P$ for 5 frequency
      bands in the Auriga region (top) and Horologium region
      (bottom). In each plot, the upper line of symbols denotes the
      tapered data, the lower line the untapered data, and the solid
      lines are linear fits to the spectra. In the untapered data, the
      spectrum is flattened at high $\ell$ due to noise.}
  \label{f9:ps_p_obs}
\end{figure}

\begin{figure}
  \centering
  \psfig{figure=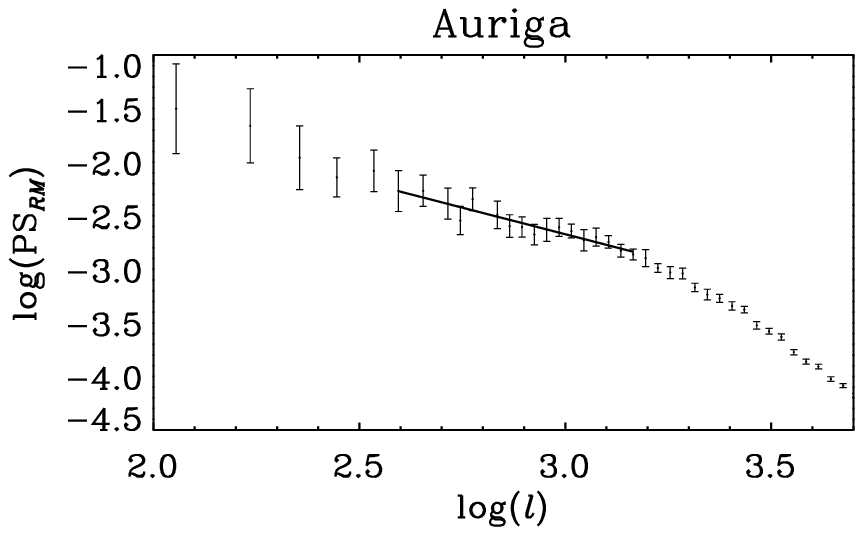,width=.4\textwidth}

  \vspace*{-0.5cm}

  \psfig{figure=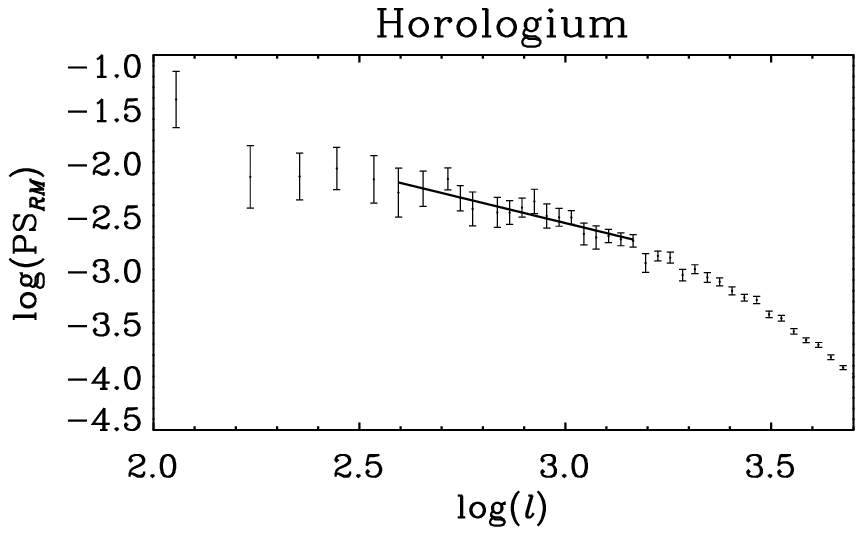,width=.4\textwidth}
  \caption{Power spectra of $RM$ in the Auriga region (top) and
      Horologium region (bottom), derived from the tapered data.}
  \label{f9:ps_rm_obs}
\end{figure}

In Fig.~\ref{f9:ps_p_obs}, we show the power spectra of $P$ in the
Auriga and Horologium regions at 5 frequencies, both for the tapered
(upper curve) and untapered data (lower curve) in the same plot. The
amplitudes of the power spectra of the tapered data are lower than
those of the untapered data only because the intensities are expressed
in mJy/beam. Because the beam widths are different for the two
datasets, this gives a difference in the magnitude of $P$ in tapered
and untapered data. The power spectra of $RM$ in the Auriga and
Horologium field are given in Fig~\ref{f9:ps_rm_obs}. Only tapered
data give reliable enough $RM$ determinations to produce power spectra
for them.

\begin{figure}
  \centering
  \psfig{figure=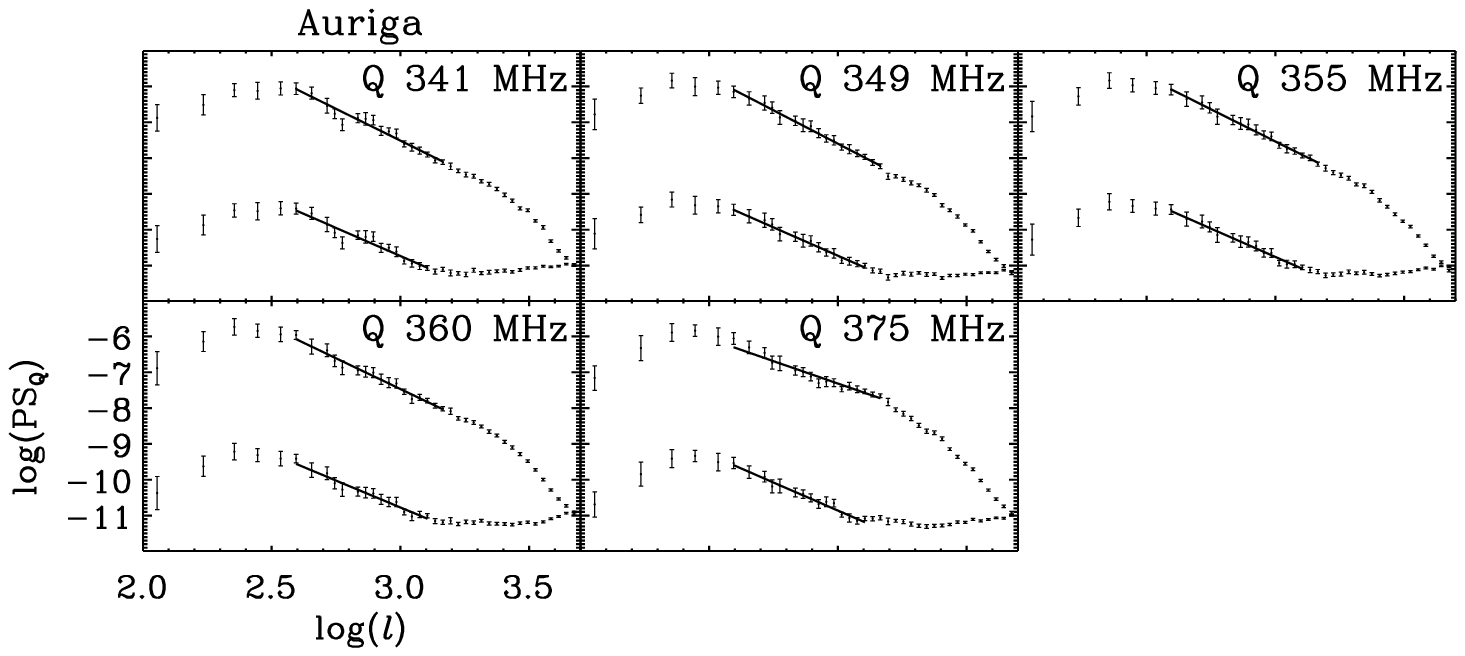,width=.5\textwidth}
  \psfig{figure=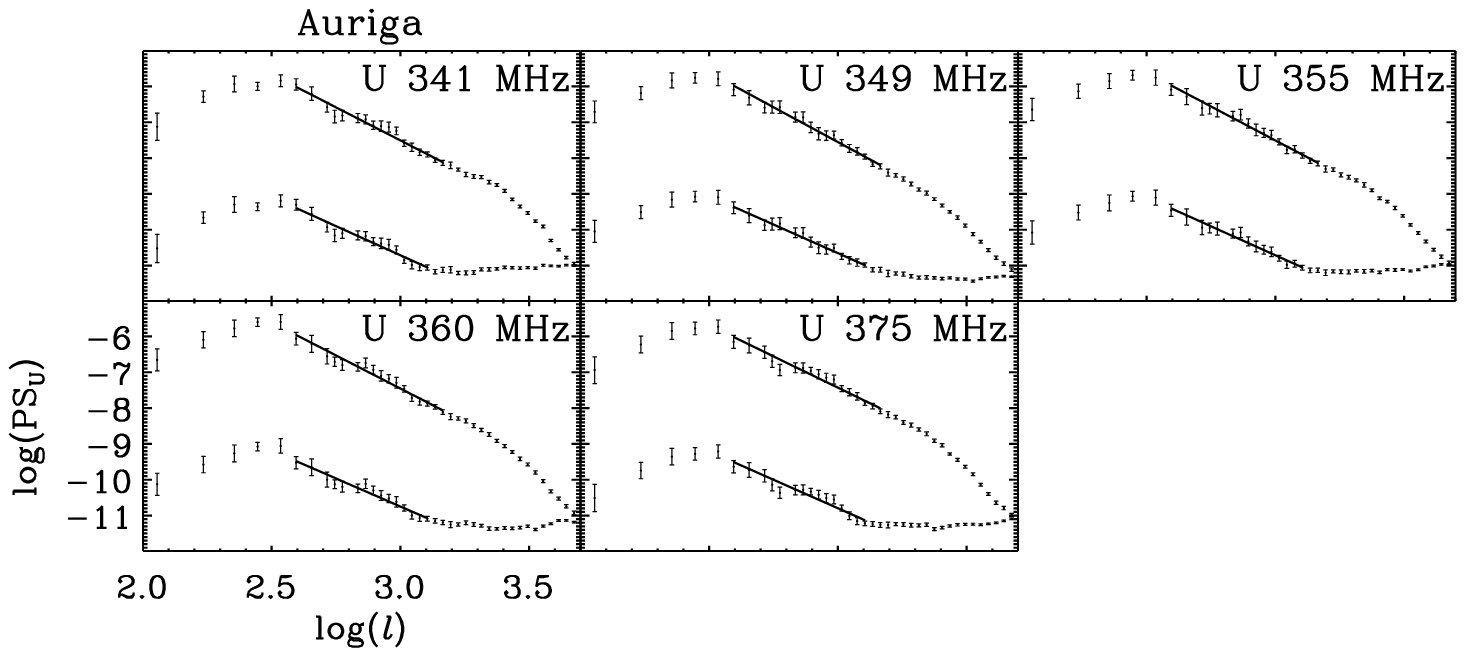,width=.5\textwidth}
  \caption{Power spectra of Stokes $Q$ (top) and Stokes $U$
      (bottom) for 5 frequencies in the Auriga region. Notation as in
      Fig.~\ref{f9:ps_p_obs}.}
  \label{f9:ps_qu_aur}
\end{figure}

\begin{figure}
  \centering
  \psfig{figure=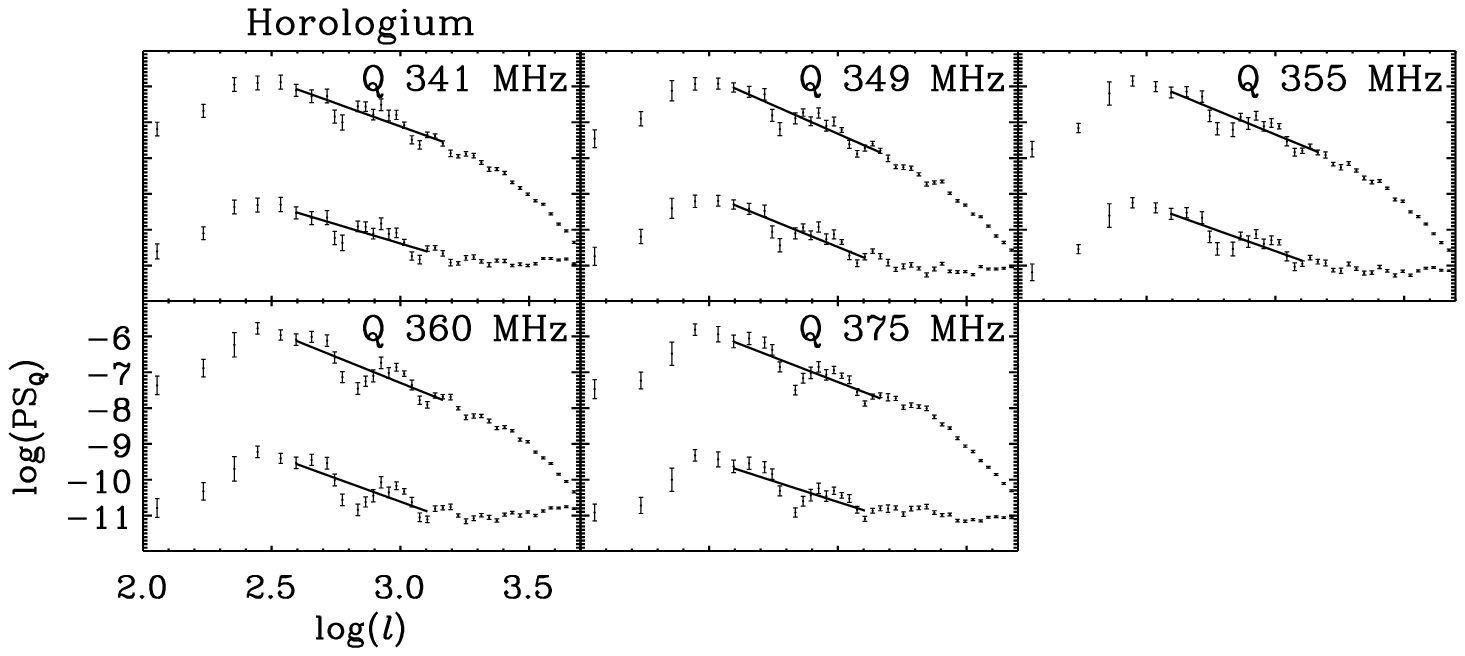,width=.5\textwidth}
  \psfig{figure=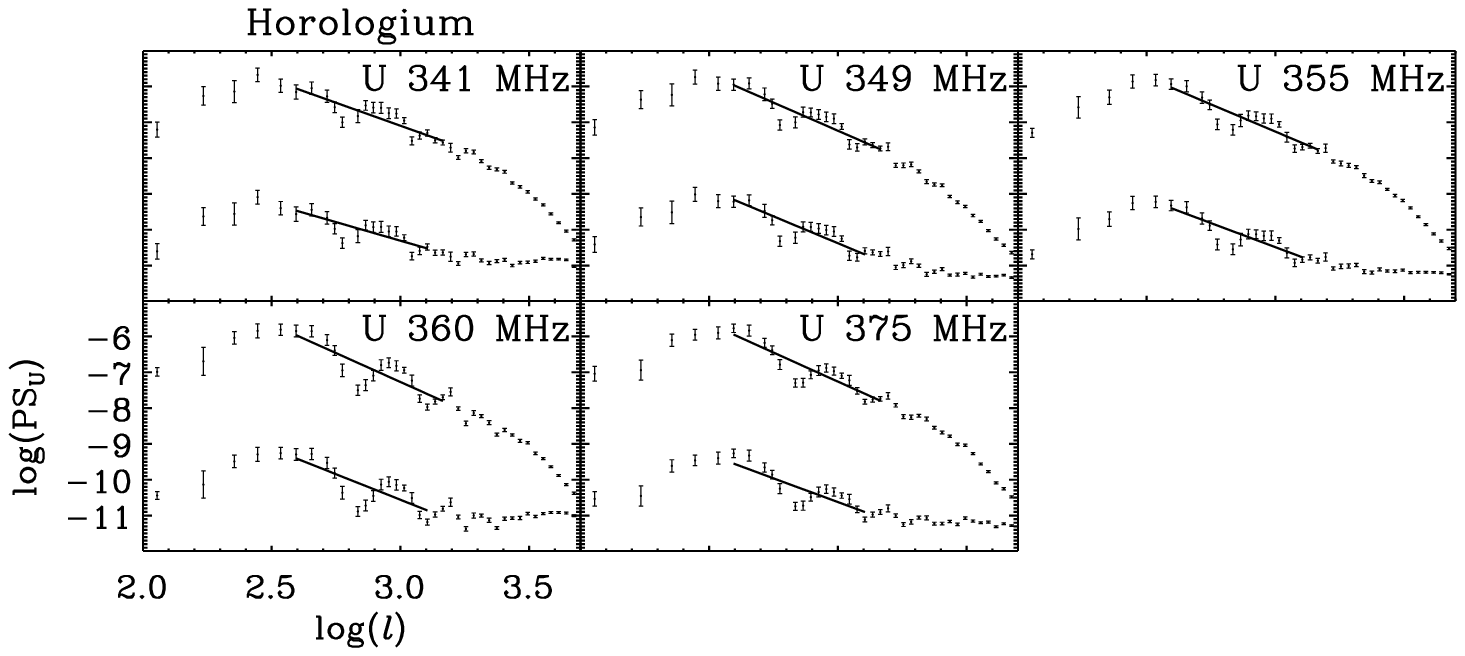,width=.5\textwidth}
  \caption{Power spectra of Stokes $Q$ (top) and Stokes $U$
      (bottom) for 5 frequencies in the Horologium region. Notation as
      in Fig.~\ref{f9:ps_p_obs}.}
  \label{f9:ps_qu_hor}
\end{figure}

Figs.~\ref{f9:ps_qu_aur} and~\ref{f9:ps_qu_hor} show power spectra for
the Stokes parameters $Q$ and $U$ in the Auriga and Horologium region
respectively, again for tapered and untapered data.  The corresponding
multipole spectral indices $\alpha$, derived for a multipole range of
$400 < \ell < 1500$, are given in Table~\ref{t9:a_obs}.

\begin{table*}
  \caption[]{Multipole spectral indices $\alpha$ for observed
  polarized intensity $P$, Stokes $Q$ and Stokes $U$, for 5
  frequencies and their average over frequency, and $\alpha_{RM}$, in
  the Auriga and Horologium regions. Values for $\alpha_P$, $\alpha_Q$
  and $\alpha_U$ are given for tapered data, denoted by a subscript
  '$t$', and untapered data. The multipole ranges used to derive
  $\alpha$ were $400 < \ell < 1500$. Only for $\alpha_P$ of the
  untapered data, the range was smaller because of flattening of the
  spectrum at higher $\ell$.}
  \vspace{0.2cm}
  \begin{center}
  \label{t9:a_obs}
  \begin{tabular}{l|r@{$\,\pm\,$}rr@{$\,\pm\,$}rr@{$\,\pm\,$}r
                  r@{$\,\pm\,$}rr@{$\,\pm\,$}r|r@{$\,\pm\,$}l}
  \multicolumn{13}{c}{\bf Auriga} \\
  \multicolumn{1}{c|}{\mbox{}}& \multicolumn{2}{c}{341 MHz} & 
  \multicolumn{2}{c}{349 MHz} & \multicolumn{2}{c}{355 MHz} & 
  \multicolumn{2}{c}{360 MHz} & \multicolumn{2}{c}{375 MHz} & 
  \multicolumn{2}{|c}{mean}\\
  \hline
  $\alpha_{P,t}$ & 2.37 & 0.19 & 2.59 & 0.20 & 2.38 & 0.19 & 
                         2.35 & 0.17 & 1.88 & 0.18 & 2.32 & 0.08 \\  
  $\alpha_P$     & 2.06 & 0.54 & 2.23 & 0.46 & 2.49 & 0.60 & 
                       2.00 & 0.47 & 2.20 & 0.59 & 2.20 & 0.24 \\  
  $\alpha_{Q,t}$ & 3.55 & 0.20 & 3.72 & 0.21 & 3.59 & 0.20 & 
                         3.43 & 0.19 & 2.50 & 0.20 & 3.36 & 0.09 \\
  $\alpha_Q$     & 3.13 & 0.24 & 3.16 & 0.25 & 3.16 & 0.24 & 
                       3.00 & 0.24 & 3.12 & 0.25 & 3.12 & 0.11\\  
  $\alpha_{U,t}$ & 3.69 & 0.19 & 3.89 & 0.21 & 3.79 & 0.21 & 
                         3.69 & 0.21 & 3.49 & 0.20 & 3.71 & 0.09 \\  
  $\alpha_U$     & 3.25 & 0.24 & 3.20 & 0.25 & 3.23 & 0.24 & 
                         3.13 & 0.24 & 3.18 & 0.53 & 3.20 & 0.11 \\   
  $\alpha_{RM,t}$ &\multicolumn{2}{c}{\mbox{}} 
  & \multicolumn{2}{c}{\mbox{}} & \multicolumn{2}{c}{\mbox{}} 
  & \multicolumn{2}{c}{\mbox{}} & \multicolumn{2}{c|}{\mbox{}} & 0.99 & 0.08\\
  \multicolumn{13}{c}{\mbox{}}\\ 
  \multicolumn{13}{c}{\bf Horologium} \\ 
  \multicolumn{1}{c|}{\mbox{} } & \multicolumn{2}{c}{341 MHz} &
  \multicolumn{2}{c}{349 MHz} & \multicolumn{2}{c}{355 MHz} &
  \multicolumn{2}{c}{360 MHz} & \multicolumn{2}{c}{375 MHz} &
  \multicolumn{2}{|c}{mean}\\ 
  \hline 
  $\alpha_{P,t}$ & 2.11 & 0.19 & 2.18 & 0.19 & 2.11 & 0.18 & 
                   1.98 & 0.19 & 1.95 & 0.19 & 2.07 & 0.08 \\ 
  $\alpha_P$     & 2.12 & 0.63 & 2.36 & 0.55 & 2.26 & 0.66 & 
		   2.47 & 0.68 & 2.49 & 0.91 & 2.34 & 0.31 \\ 
  $\alpha_{Q,t}$ & 2.58 & 0.22 & 3.17 & 0.19 & 2.96 & 0.19 & 
                   2.90 & 0.20 & 2.78 & 0.20 & 2.89 & 0.09 \\
  $\alpha_Q$     & 2.15 & 0.26 & 2.92 & 0.22 & 2.57 & 0.23 & 
		   2.60 & 0.23 & 2.32 & 0.25 & 2.51 & 0.11 \\ 
  $\alpha_{U,t}$ & 2.56 & 0.22 & 3.14 & 0.20 & 3.04 & 0.19 & 
                   3.22 & 0.20 & 3.24 & 0.17 & 3.04 & 0.09 \\ 
  $\alpha_U$     & 2.07 & 0.26 & 3.00 & 0.22 & 2.69 & 0.23 & 
		   2.85 & 0.23 & 2.65 & 0.25 & 2.65 & 0.11 \\ 
  $\alpha_{RM,t}$ &\multicolumn{2}{c}{\mbox{}} & 
  \multicolumn{2}{c}{\mbox{}} & \multicolumn{2}{c}{\mbox{}} & 
  \multicolumn{2}{c}{\mbox{}} & \multicolumn{2}{c|}{\mbox{}} & 0.94 & 0.10\\
    \end{tabular}
  \end{center}
\end{table*}

At small scales (large $\ell$), the power spectra of the untapered
data flatten out due to the noise in the maps, while the
low-resolution data steepen due to the tapering, as illustrated in
Fig.~\ref{f9:ps_corr}.  At large scales, the $Q$ and $U$ power spectra
of the tapered data show a decrease. This decrease could be due to the
lack of large-scale structure (see Sect.~\ref{s9:wsrtdata}), but then
it is hard to explain why there is no such decline in $P$.

The power spectra of $Q$ and $U$ are steeper and have a larger
amplitude than the power spectra of $P$. This could be caused by the
presence of a Faraday screen in front of the emitting region. A
Faraday screen will rotate the polarization angle, and so induce extra
structure in $Q$ and $U$, while leaving $P$ unaltered. This results in
a higher amplitude of the power spectrum. As the Faraday screen
consists of foreground material, its angular size is large, steepening
the spectrum. This effect was also noticed by Tucci et al.\ (2002).

The logarithmic slope of the power spectrum of polarized intensity is
$\alpha_P \approx$ 2.1 - 2.3 (Table~\ref{t9:a_obs}), which is slightly
higher than most earlier estimates from the literature (although those
are taken at higher frequencies, see Sect.~\ref{s9:lit_ps}).
However, note that the Auriga and Horologium regions were selected for
their conspicuous structure in $P$, so we expect these regions to show
more structure on large (degree) scales than the ``average'' ISM, and
thus exhibit a steeper spectrum.  The power spectra of $Q$ and $U$ in
the Auriga region are somewhat steeper than in Horologium,
indicating that the Horologium region probably contains more
small-scale structure in the Faraday screen than the Auriga region.

The power spectra of $RM$ in Fig.~\ref{f9:ps_p_obs} are shallower than
the $Q$, $U$ or $P$ power spectra. In fact, we do not expect a direct
correspondence between the multipole spectral indices of $RM$ and $P$
(or $Q$, $U$), as the former describes very directly the electron
content and magnetic field in the ISM (integrated over the line of
sight), whereas in the latter case the polarized radiation is
modulated by Faraday rotation and depolarization.

\subsection{Power spectra from the WENSS polarization region}
\label{ss9:psw}

\begin{table}
  \caption[]{Multipole spectral indices $\alpha$ for observed
  polarized intensity $P$, Stokes $Q$ and Stokes $U$ in 11 subfields
  in the WENSS polarization region at 325~MHz.}
  \vspace{0.2cm}
  \begin{center}
  \label{t9:a_wenss}
  \begin{tabular}{l|c|r@{$\,\pm\,$}lr@{$\,\pm\,$}lr@{$\,\pm\,$}l}
  No. & $(l,b) \;(\dg,\dg)$ & \multicolumn{2}{c}{$\alpha_P$}
  & \multicolumn{2}{c}{$\alpha_Q$} & \multicolumn{2}{c}{$\alpha_U$} \\
  \hline
  1  & (159, 04) & 1.67 & 0.08 & 2.55 & 0.09 & 2.43 & 0.10 \\
  2  & (165, 11) & 1.48 & 0.08 & 2.10 & 0.08 & 2.06 & 0.08 \\
  3  & (158, 11) & 1.84 & 0.09 & 2.50 & 0.09 & 2.20 & 0.09 \\
  4  & (151, 11) & 1.70 & 0.08 & 2.09 & 0.09 & 2.02 & 0.09 \\
  5  & (144, 11) & 1.63 & 0.08 & 2.28 & 0.10 & 2.31 & 0.10 \\
  6  & (143, 18) & 1.24 & 0.09 & 1.96 & 0.10 & 2.04 & 0.10 \\
  7  & (150, 18) & 0.99 & 0.08 & 1.44 & 0.08 & 1.47 & 0.09 \\
  8  & (157, 18) & 1.67 & 0.09 & 2.33 & 0.09 & 2.33 & 0.09 \\
  9  & (157, 25) & 0.73 & 0.08 & 1.24 & 0.08 & 1.18 & 0.09 \\
  10 & (150, 25) & 0.90 & 0.08 & 1.18 & 0.08 & 1.44 & 0.08 \\
  11 & (144, 25) & 0.68 & 0.08 & 0.77 & 0.08 & 0.63 & 0.08 \\
  \end{tabular}
  \end{center}
\end{table}

\begin{figure}
  \centering
  \psfig{figure=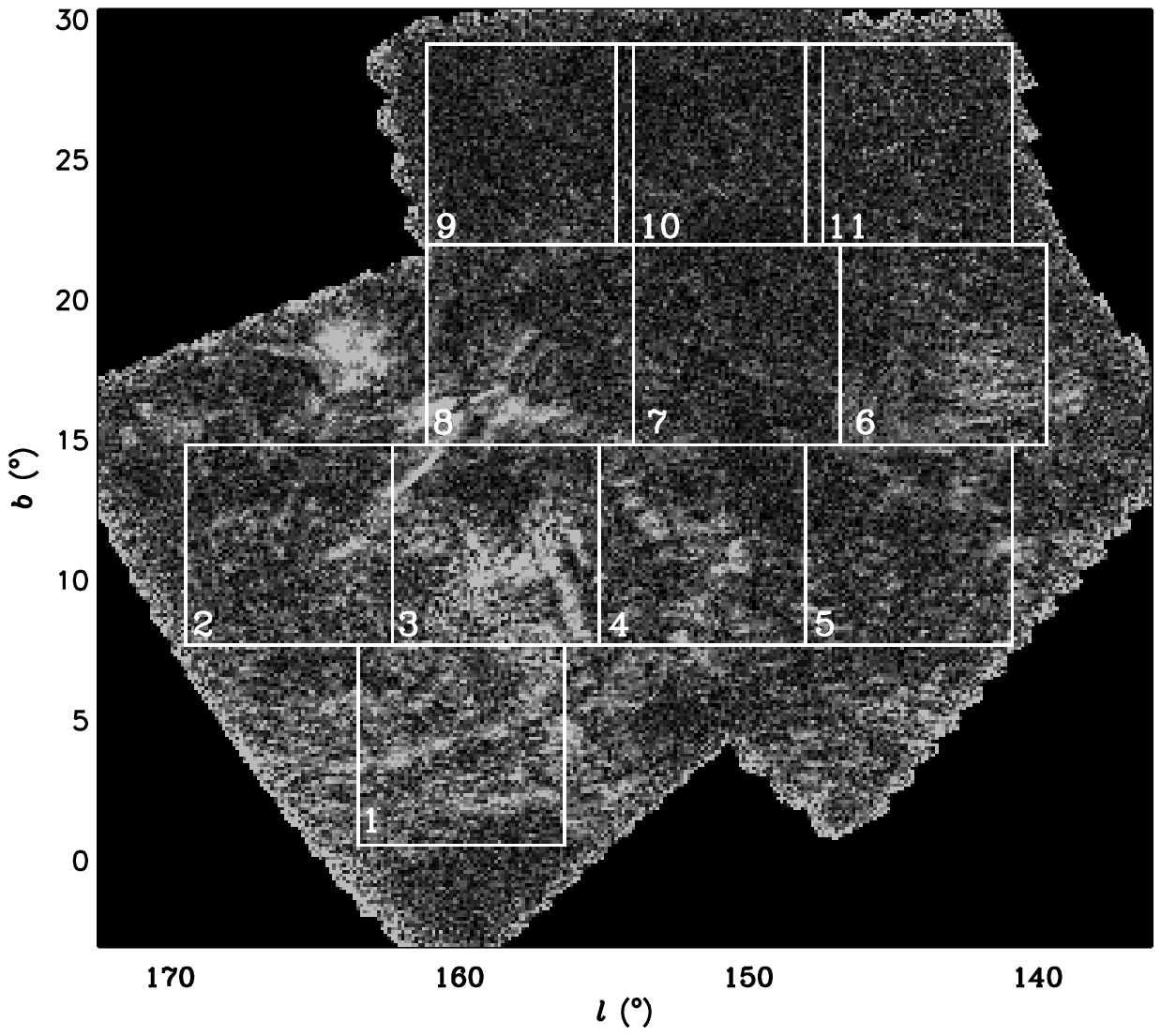,width=0.5\textwidth}
  \caption{Polarized intensity at 325~MHz of the WENSS region in
      grey scale, superimposed with white boxes denoting the numbered
      subfields over which angular power spectra were computed.}
  \label{f9:wenssboxes}
\end{figure}

\begin{figure}
  \centering
  \psfig{figure=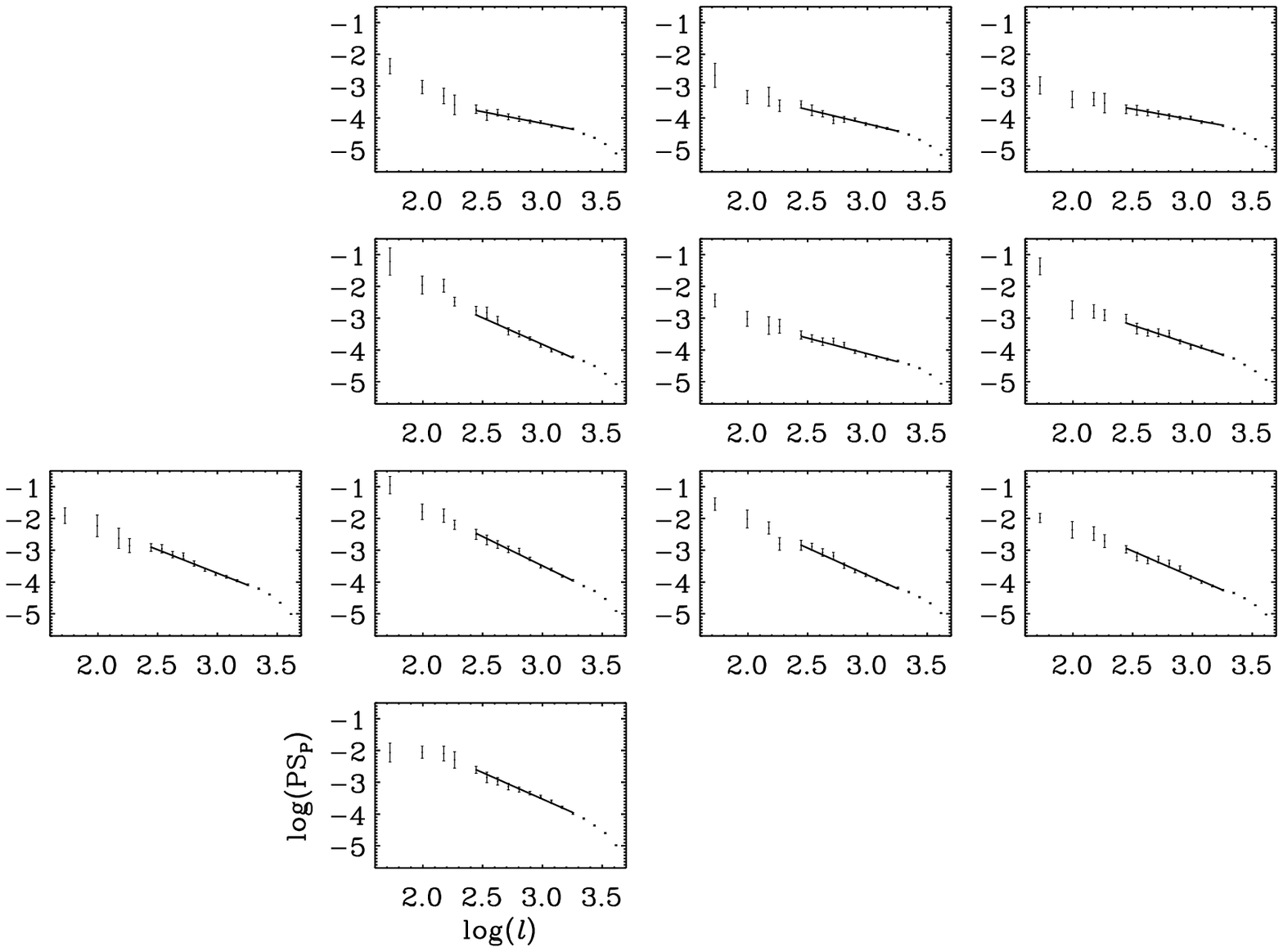,width=.5\textwidth}
  \caption{Power spectra of polarized intensity $P$ in subfields
      in the WENSS polarization region. The plots are arranged as in
      Fig.~\ref{f9:wenssboxes}.}
  \label{f9:ps_wr_pi}
\end{figure}

In the WENSS polarization region, power spectra were evaluated for
subfields, to study possible dependences of the multipole
spectral index on Galactic longitude and/or latitude. The 11 subfields
are shown in Fig.~\ref{f9:wenssboxes}, superimposed on grey scale maps
of $P$. The power spectra of polarized intensity $P$ are shown in
Fig.~\ref{f9:ps_wr_pi}, where the subfields are arranged as in
Fig.~\ref{f9:wenssboxes}. The power spectra in subfields 9, 10 and 11,
at high Galactic latitude $b$, have a lower amplitude than the power
spectra at lower $b$, which is consistent with the decreasing amount
of $P$ at higher $b$ visible in Fig.~\ref{f9:wenss}.

\begin{figure}
  \centering
  \psfig{figure=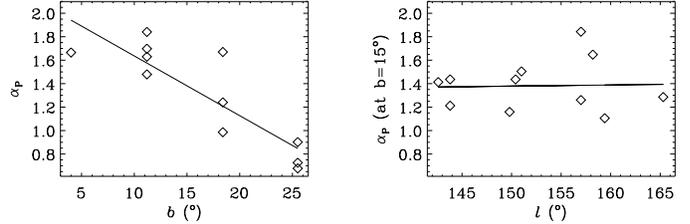,width=.5\textwidth}
  \caption{Dependence of multipole spectral index $\alpha_P$ on
      Galactic latitude $b$ (left) and longitude $l$ (right), for 11
      subfields in the polarization part of the WENSS survey. The
      spectral indices in the right plot are normalized to the
      latitude gradient shown in the left plot.}
  \label{f9:ps_wr_lb}
\end{figure}

The multipole spectral indices of the power spectra of $P$, $Q$ and
$U$ are given in Table~\ref{t9:a_wenss}, and the dependence of
$\alpha_P$ on Galactic longitude and latitude is shown in
Fig.~\ref{f9:ps_wr_lb}. The observed decrease of spectral index with
increasing latitude (i.e.\ power spectra become flatter with
increasing latitude) indicates a decrease in the amount of large-scale
structure with increasing latitude. The dependence of spectral index
on longitude is computed after rescaling of the data to a standard
latitude of $b = 15$\dg.  The fitted slopes in Fig.~\ref{f9:ps_wr_lb},
and slopes determined for $\alpha_Q$ and $\alpha_U$ in a similar way,
are: 
\[
\left\{ 
\begin{array}{ll}
  \partial\alpha_P/\partial b = -0.051 \pm 0.004 \hspace*{.3cm} &
  \partial\alpha_P/\partial l = -0.001 \pm 0.003 \\ 
  \partial\alpha_Q/\partial b = -0.073 \pm 0.004 &
  \partial\alpha_Q/\partial l = -0.008 \pm 0.004 \\
  \partial\alpha_U/\partial b = -0.066 \pm 0.004 &
  \partial\alpha_U/\partial l = -0.005 \pm 0.004 \\
\end{array} \right.
\]
In summary, the spectral index $\alpha$ in $P$, $Q$ and $U$ decreases
with increasing Galactic latitude, and is consistent with no
dependence of $\alpha$ on Galactic longitude. Although the small
errors in the derived slopes suggest a good determination of the
slope, the large spread of the data points in Fig.~\ref{f9:ps_wr_lb}
indicates that a linear gradient is not the perfect model to describe
the data.

The decrease of $\alpha$ with latitude indicates that there is more
structure in polarization on larger scales at lower latitude. Thus,
although small-scale structure in polarization is seen up to very high
latitudes, even at frequencies as low as 350~MHz (Katgert \& de Bruyn
1999), the amount of structure and its spectral index decrease with
latitude, at least in this region of the sky. The region studied here
may be special, as it is situated at the edge of a region of high
polarization (the ``fan region'', see e.g. Brouw \& Spoelstra 1976),
which is thought to have little structure in magnetic field and/or
electron density compared to its surroundings. This high-polarization
region extends to $b \approx 20$\dg, and its edge could be responsible
for the decrease in structure on scales of $\sim$~1\dg.

The spectral indices in the Auriga ($b = 16\dg$) and Horologium ($b =
7\dg$) regions are higher than implied by Fig.~\ref{f9:ps_wr_lb}. This
might be due to the way in which the Auriga and Horologium regions
were selected, viz.\ because of their remarkable structure on degree
scales.

\subsection{Existing literature of power spectra from diffuse polarization}
\label{s9:lit_ps}

Much work has been done on the determination of power spectra of the
diffuse Galactic synchrotron background, because the Galactic
synchrotron radiation is a foreground contaminator in Cosmic Microwave
Background Radiation (CMBR) polarization measurements at high
frequencies $\nu \approx$ 30 - 100~GHz. Power spectra of the diffuse
polarized synchrotron background intensity have been determined from several
radio surveys at frequencies from 408 MHz to 2.7 GHz, in many parts of
the sky (Tucci et al.\ 2000, 2002, Baccigalupi et al.\ 2001,
Giardino et al.\ 2002, Bruscoli et al.\ 2002).

These power spectra studies are based on the following surveys of
polarized radiation:
\begin{itemize}
\item Dwingeloo 25m-dish survey (Brouw \& Spoelstra 1976), of the
      region $120\dg <l<180\dg$ and $b > -10$\dg (although
      undersampled). This is a multi-frequency survey at 408~MHz,
      465~MHz, 610~MHz, 820~MHz and 1411~MHz, with increasing angular
      resolution of 2.3\dg\ to 0.5\dg.
\item Parkes 2.4~GHz Galactic plane survey (Duncan et al.\ 1997), of
      the region $238\dg <l<5\dg$ and with $|b| < 5$\dg, at some
      positions a few degrees higher, at a resolution of 10.4\arcmin.
\item Effelsberg 2.695~GHz Galactic plane survey (Duncan et al.\
      1999), of the region $5\dg <l<74\dg$, and $|b| < 5\dg$, at a
      resolution of 4.3\arcmin.
\item Effelsberg 1.4~GHz intermediate latitude survey (Uyan\i ker et 
      al.\ 1999), which consists of 4 regions within $45\dg
      <l<210\dg$ and $-15\dg <b<20\dg$ with an angular resolution of
      10.4\arcmin.
\item Australia Telescope Compact Array (ATCA) 1.4~GHz survey 
      (Gaensler et al.\ 2001). This is a test region for the Southern
      Galactic Plane Survey (SGPS, McClure-Griffiths et al.\ 2001) at
      $325.5\dg < l < 332.5\dg$, $-0.5\dg < b < 3.5\dg$, with a
      resolution of about 1\arcmin.
\end{itemize}
Power spectra of total intensity $I$ and polarized intensity $P$ were
derived in these surveys for multipoles over a range of $\ell \approx
10$ to 6000. 

\begin{figure*}
  \centering 
  \psfig{figure=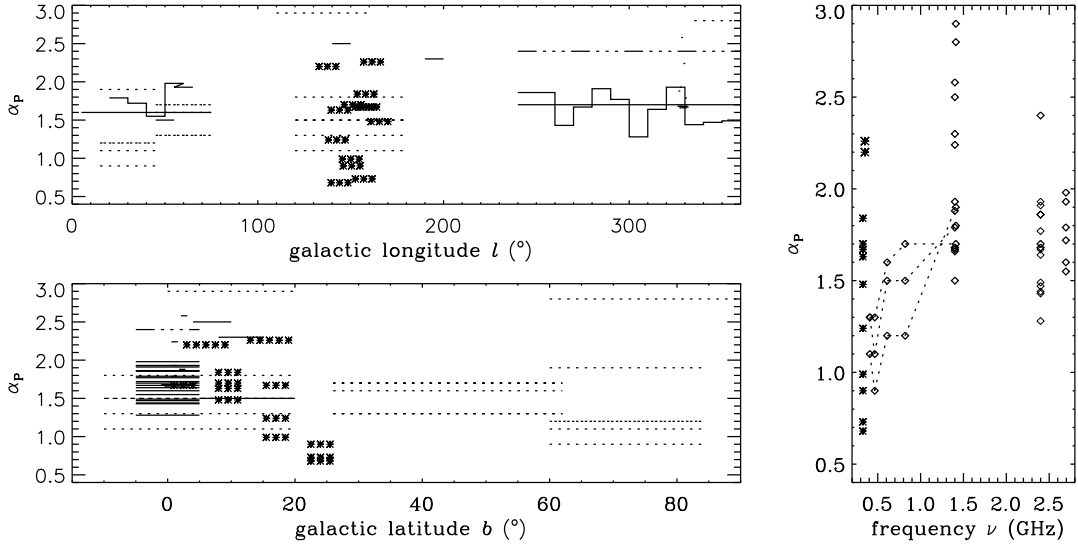,width=.8\textwidth}
  \caption{Estimates of multipole spectral indices $\alpha_P(\ell)$ from
         our observations and from the literature, as a function of
         Galactic longitude (top), Galactic latitude (bottom), and
         frequency (right). In the left-hand plots, solid lines denote
         high multipole numbers ($100 < \ell < 6000$), the dashed-dotted
         line is an intermediate multipole range ($30 < \ell < 200$)
         and the dotted lines give small multipoles ($10 < \ell <
         80$). In the right-hand (frequency) plot, the dotted lines
         connect observations of the same region made at different
         frequencies. The WSRT observations discussed in this paper
         are shown by asterisks.}
  \label{f9:lit}
\end{figure*}

Fig.~\ref{f9:lit} shows the variation of $\alpha_P$ with Galactic
longitude, latitude and frequency, using the available data as
detailed in Table~\ref{t9:lit}. In the left plots, the lines show
ranges in longitude (top) and latitude (bottom) over which $\alpha_P$
was computed. Solid lines give high multipole numbers ($100 < \ell <
6000$), dashed-dotted lines denote an intermediate multipole range
($30 < \ell < 200$) and the dotted lines give small multipoles ($10 <
\ell < 80$). The WSRT data from the Auriga, Horologium en WENSS
fields, discussed here, are given in asterisks. In the right plot,
$\alpha_P$ against frequency is displayed. The connected points are
from the same area observed at different frequencies from 408~MHz to
1411~MHz (in the Brouw \& Spoelstra paper).

The first conclusion from Fig.~\ref{f9:lit} is that multipole spectral
indices vary over the sky from $\alpha \approx 1$ to 3, without
showing a clear correlation with Galactic longitude, while only the
WENSS subfields show a dependence of spectral index on latitude.
However, all surveys have been done at different resolutions and
frequencies, and the regions used to compute power spectra are of
different sizes. This can explain why a possible dependence of
$\alpha$ on latitude was not clearly seen in the other studies. The
large variation in slopes of angular power spectra in $P$ indicates
that interpretation of the slope is not straightforward, possibly due
to large influence of depolarization mechanisms. Care must therefore
be taken in extrapolating the results to higher frequencies.

Furthermore, spectral indices show an increase with frequency from
408~MHz to 1.4~GHz. This means that the power spectra become steeper,
so the relative amount of small-scale structure decreases. This could
be due to the large Faraday rotation at low frequencies. Typical $RM$s
of 5~\radm\ are present in the Brouw \& Spoelstra data (Spoelstra
1984), and will rotate polarization angles at 325~MHz by about
250\dg. Variations in $RM$ of a few \radm\ give angle variations of
over 90\dg, which would cause beam depolarization if the angle
variations occur on scales smaller than the beam (2.3\dg\ in this
case). Beam depolarization only acts on scales of the synthesized
beam, and therefore creates additional structure on small scales in
$P$, which flattens the power spectrum. A $\Delta RM$ of 5~\radm\
would cause a variation in polarization angle of about 40\dg\ at
820~MHz, and of no more than 10\dg\ at 1.4~GHz. So at frequencies
above 1.4~GHz, a $\Delta RM$ of 5~\radm\ would cause negligible beam
depolarization. In addition, the resolution of the observations
generally increases with increasing frequency, which would also cause
a decrease in beam depolarization. This might explain why above
1.4~GHz the spectral index does not appear to be correlated with
frequency.  The fact that spectral indices in the two WSRT regions are
much higher than would be expected from this argument can be due to
the criteria used to select the two fields.

\section{Structure functions}
\label{s9:sf}

The disadvantage of using angular power spectra is that a regular grid
of data is required. If the data are very irregularly spaced (e.g.\ in
the case of data from pulsars or extragalactic point sources), it is better
to use the structure function which in principle gives the same
information, but can be calculated easily for irregularly spaced
data. The structure function $SF$ of a radiation field $X$, as a
function of distance lag $d$, is
\begin{equation}
  SF_X(d) = \frac{\sum_{i=1}^N (X(x_i) - X(x_i+d))^2}{N}
  \label{e:sf}
\end{equation}
where $X(x_i)$ is the value of field $X$ at position $x_i$ and $N$ is
the number of data points. If the power spectrum of $RM$ is a power
law with spectral index $\alpha$, then the structure function
$SF_{RM}$ is
\begin{equation}
  SF_{RM}(d) \propto d^{\,\mu} \hspace*{0.3cm} \mbox{where} \;\;\;
  \mu        = \left\{ \begin{array}{ll}
                  \alpha - 2 & \;\;\mbox{for } 2 < \alpha < 4 \\
		  2          & \;\;\mbox{for } \alpha > 4 \\
		\end{array}
		\right.
  \label{e9:sf}
\end{equation}
(Simonetti et al.\ 1984). We determined the structure functions of
$RM$ to compare with existing estimates of the structure function of
Galactic $RM$ from polarized extragalactic point sources. As the
determination of structure functions does not require a regular grid,
we can compute the $SF$ including and excluding ``bad'' data points to
examine how the structure functions of $P$ and $RM$ change. This will
allow us to estimate the effect of ``bad'' data in the power spectra
in $P$ and $RM$ computed in Sect.~\ref{s9:ps}.

\subsection{Structure functions of $RM$}

\begin{figure}
  \centering
  \psfig{figure=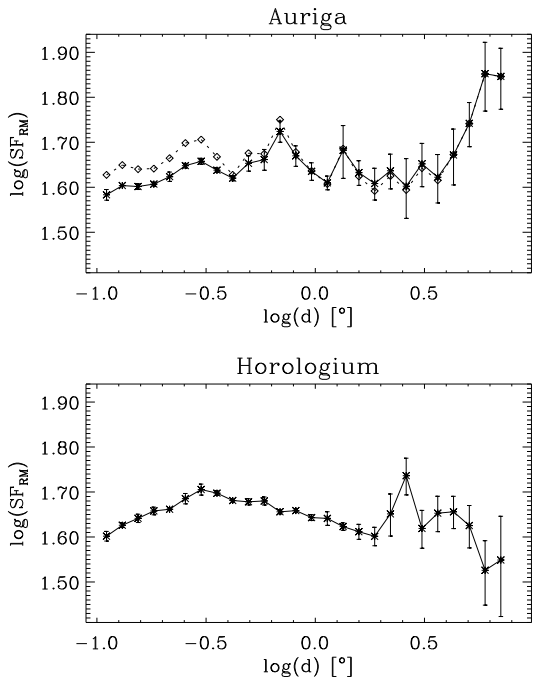,width=.4\textwidth}
  \caption{The solid lines show the $RM$ structure function
         $SF_{\scriptsize RM}$ as a function of distance $d$ in
         degrees for the Auriga region (top) and the Horologium
         region (bottom), using only reliable $RM$ values. The dotted
         line in the top plot gives $SF_{\scriptsize RM}$ evaluated
         for the entire grid, including unreliable $RM$
         determinations.}
  \label{f9:sf_rm}
\end{figure}

Structure functions $SF_{\scriptsize RM}$ in the Auriga and Horologium
regions are plotted against distance $d$ in degrees in the log-log
plots in Fig.~\ref{f9:sf_rm} (solid line), where error bars denote the
standard deviation. The minimum distance shown is $d
\approx$~7\arcmin.  For the evaluation of the SF, only ``reliably
determined'' $RM$ values are used, according to the definition in
Sect.~\ref{s9:wsrtdata}.  Although the spectrum is consistent with a
flat slope, there is some evidence for a break in the slope at $d =
0.3\dg$, primarily in the Horologium field. For larger angular scales,
the SF is approximately flat in the Auriga region, with a tentative
increase at the largest lags, and even decreasing in Horologium. 

We can estimate the magnitude of the contribution of unreliably
determined $RM$s by reevaluating the SF for the complete grid of $RM$
values, instead of only the reliable $RM$s. This estimate is important
for a discussion of the power spectra of $RM$, which were evaluated
over the complete dataset, including unreliably determined $RM$s.  The
structure function using the complete dataset is shown in the left
panel of Fig.~\ref{f9:sf_rm} as a dotted line. The structure function
clearly has a lower amplitude if the unreliable $RM$ determinations
are removed, but the slope of the structure function remains
approximately the same.

\subsection{Structure functions of $P$}

\begin{figure*}
  \centering
  \psfig{figure=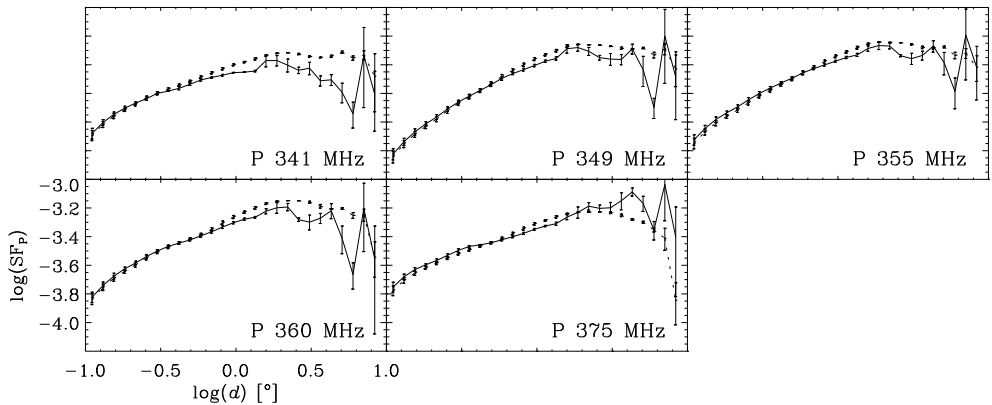,width=.75\textwidth} 
  \caption{Structure function $SF_P$ as a function of distance $d$ in
           degrees for the Auriga region, in 5 frequencies. The solid
           line shows $SF_P$ in which only beams with $P > 4\sigma$
           are included, the dotted line gives $SF_P$ as computed from
           all data.}
  \label{f9:sf_p}
\end{figure*}

We compute the structure function of polarized intensity, for both the
complete grid of beams, and for those beams selected to have high
$P$. In Fig.~\ref{f9:sf_p} we show structure functions of $P$ in the
Auriga region for 5 frequencies, where again the error bars
  denote the standard deviation. The average logarithmic slope of the
structure functions is $\sim$~0.35 in the range $0.2\dg \la d \la
1\dg$, and the spectrum flattens on larger scales, probably due to the
insensitivity of the WSRT to structure on large angular scales. The
solid line denotes the estimate based  only on beams with $P >
4\sigma$, and the dotted line gives the estimate based on all the data.
The structure functions based on selected beams and those based on all
data are not significantly different on small scales, and start
deviating only for $d \ga 0.7$\dg. Although small deviations are created by
selecting the best data, the overall slope of the structure function
is hardly affected by the selection including only regions of high
$P$.  The relatively small effect of including bad data gives
confidence that the determination of the logarithmic slope of the {\it
power spectra} of $RM$ and $P$ is not significantly influenced by
noisy data or poorly determined $RM$.

\section{Discussion}
\label{s9:disc}

To the authors knowledge, the results presented here are the first
observational determinations of the power spectrum of rotation
measure.  What power spectrum is to be expected is unclear, as the
$RM$ is a complex quantity depending on magnetic field structure and
direction, electron density and the path length through the ionized
ISM. There are indications that the electron density and $RM$ exhibit
Kolmogorov turbulence (Minter \& Spangler 1996, Armstrong et al.\ 1995).

An approximately flat structure function corresponds to structure of
equal amplitude on all scales, similar to a noise spectrum. The break
in $SF_{RM}$ at $\delta\theta \approx 0.3\dg$, if present, corresponds
to a change in characteristics of the structure on scales of
$\sim$~3.9~pc, assuming a path length of 600~pc (Haverkorn et al.\
2003d).

Minter \& Spangler (1996) studied structure functions of $RM$ from
polarized extragalactic sources and of emission measure ($EM$) from
H$\alpha$ measurements, on the same angular scales as in our
observations. They find a break in the slope of the structure
function, which can be interpreted as a transition of 3D Kolmogorov
turbulence ($\mu = 5/3$, where $\mu$ is the slope of the structure
function as defined in Eq.~(\ref{e9:sf})) to 2D turbulence ($\mu =
2/3$) in $RM$ on angular scales of $\delta\theta \approx 0.1\dg$, the
scale of our resolution.  They assume a total path length of 2000~pc,
and conclude that the transition occurs at scales of $\sim$~3.6~pc,
consistent with the spatial scale at which we estimate the turnover
point in the SF of the Auriga and Horologium regions. However, there
is a large discrepancy between $\mu$ values found by Minter \&
Spangler ($\mu = 5/3$ and $2/3$) and by us ($\mu \approx 0$), which
could be explained by the fact that they probe the complete line of
sight through the medium of many kpc, whereas the $RM$ that we obtain
is only produced in the nearest $\sim$~600~pc.  Therefore, it is
possible that the nature of the turbulence changes from $\mu = 0$
nearby (approximately in the Galactic stellar disk), to 2D and/or
Kolmogorov-like turbulence at larger distances.

However, flat structure functions are also found by Simonetti et al.\
(1984), who have studied structure functions from $RM$s in three
regions of the sky. Two of the three regions (around the North
Galactic Pole and at $180\dg < l< 220\dg$, $10\dg < b < 50\dg$) are
consistent with a flat structure function ($\mu = 0$). (The third
region is located along the Local arm and in the plane, at
$70\dg<l<110\dg$ and $-45\dg<b<5$, and its structure function of $RM$
has a slope $\mu \approx 1$.) The smallest angular scale they probe in
these regions is about 2\dg.  Clegg et al.\ (1992) study low-latitude
extragalactic sources and also obtain flat structure functions, albeit
with a much higher amplitude than that of the high-latitude
extragalactic sources.

\section{Conclusions}
\label{s9:conc}

The multipole spectral index for polarized intensity $P$ is $\alpha
\approx 2.2$ in the Auriga and Horologium regions, and ranges from 0.7
to 1.8 for subfields in the WENSS polarization region. The multipole
spectral index decreases with Galactic latitude (i.e.\ power spectra
become flatter towards higher latitudes), but is probably constant
with Galactic longitude.

In all regions, the power spectra of Stokes $Q$ and $U$ are steeper
than the spectra of $P$. This is most likely due to the presence of a
Faraday screen, which creates additional structure in $Q$ and $U$, but
not in $P$. As the Faraday screen is located in front of the polarized
emission, the structure induced by the screen will be on larger
angular scales than that of the emission, which steepens the $Q$ and $U$
spectra. The derived power spectra of $P$ agree with earlier estimates,
although all estimates show a large range of $0.6 \la \alpha_P \la 3$,
possibly due to a large influence of depolarization mechanisms. This
makes interpretation of the power spectra uncertain.

Structure functions of $RM$ in the Auriga and Horologium fields are
consistent with flat spectra (i.e.\ a logarithmic slope of the
structure function $\mu = 0$), but may show a break close to 0.3\dg,
which is at the same spatial scales as a break in the structure
function in the $RM$s of extragalactic sources (Minter \& Spangler
1996). The flat spectrum indicates a noise-like spectrum with equal
amounts of structure on all scales.

The derived structure functions support the estimates of the power
spectra, because they are based only on high-quality data,
i.e. reliably determined $RM$, and $P$ with high S/N. Structure
functions using all data do not show a significant difference with the
structure functions of the selected data. This gives confidence that
the power spectra determinations are not significantly affected by bad
data points.

\section*{Acknowledgements}
We thank B. J. Rickett for helpful discussions and suggestions, and
F. Heitsch for help in constructing and analyzing power spectra.
The Westerbork Synthesis Radio Telescope is operated by the
Netherlands Foundation for Research in Astronomy (ASTRON) with
financial support from the Netherlands Organization for Scientific
Research (NWO).  Computations presented here were performed on the SGI
Origin 2000 machine of the Rechenzentrum Garching of the
Max-Planck-Gesellschaft. MH acknowledges the support from NWO grant
614-21-006.

\begin{table*}[h!]
  \centering
  \begin{tabular}{lr@{ - }lr@{ - }lrr@{ - }lrl} 
  survey & \multicolumn{2}{c}{$l\;(\dg)$} &
  \multicolumn{2}{c}{$b\;(\dg)$} & $\nu$ (MHz) &
  \multicolumn{2}{c}{$\ell$} & $\alpha_P$ & reference \\
  \hline\hline
Dwingeloo         & 110 & 160                   &    0 & 20                   & 1411 &  30 & 100 & 2.9  & Baccigalupi et al.\ 2001\\
Brouw \& Spoelstra&   5 &  80                   &   50 & 90                   &  ''  &\multicolumn{2}{c}{\mbox{''}}& 3.1 &\\
(1976)            & 335 & 360                   &   60 & 90                   &  ''  &\multicolumn{2}{c}{\mbox{''}}& 2.8 &\\
                  & 120 & 180                   & --10 & 20                   &  408 &  10 &  70 & 1.3  & Bruscoli et al.\ 2002\\
                  &\multicolumn{2}{c}{\mbox{''}}&\multicolumn{2}{c}{\mbox{''}}&  465 &\multicolumn{2}{c}{\mbox{''}}& 1.1 &\\
                  &\multicolumn{2}{c}{\mbox{''}}&\multicolumn{2}{c}{\mbox{''}}&  610 &\multicolumn{2}{c}{\mbox{''}}& 1.5 &\\
                  &\multicolumn{2}{c}{\mbox{''}}&\multicolumn{2}{c}{\mbox{''}}&  820 &\multicolumn{2}{c}{\mbox{''}}& 1.5 &\\
                  &\multicolumn{2}{c}{\mbox{''}}&\multicolumn{2}{c}{\mbox{''}}& 1411 &\multicolumn{2}{c}{\mbox{''}}& 1.8 &\\
                  &  45 &  75                   &   26 & 62                   &  408 &  10 &  70 & 1.3  & \\                
                  &\multicolumn{2}{c}{\mbox{''}}&\multicolumn{2}{c}{\mbox{''}}&  465 &\multicolumn{2}{c}{\mbox{''}}& 1.3 &\\
                  &\multicolumn{2}{c}{\mbox{''}}&\multicolumn{2}{c}{\mbox{''}}&  610 &\multicolumn{2}{c}{\mbox{''}}& 1.6 &\\
                  &\multicolumn{2}{c}{\mbox{''}}&\multicolumn{2}{c}{\mbox{''}}&  820 &\multicolumn{2}{c}{\mbox{''}}& 1.7 &\\
                  &\multicolumn{2}{c}{\mbox{''}}&\multicolumn{2}{c}{\mbox{''}}& 1411 &\multicolumn{2}{c}{\mbox{''}}& 1.7 &\\
                  &  15 &  45                   &   60 & 84                   &  408 &  10 &  70 & 1.1  & \\       
                  &\multicolumn{2}{c}{\mbox{''}}&\multicolumn{2}{c}{\mbox{''}}&  465 &\multicolumn{2}{c}{\mbox{''}}& 0.9 &\\
                  &\multicolumn{2}{c}{\mbox{''}}&\multicolumn{2}{c}{\mbox{''}}&  610 &\multicolumn{2}{c}{\mbox{''}}& 1.2 &\\
                  &\multicolumn{2}{c}{\mbox{''}}&\multicolumn{2}{c}{\mbox{''}}&  820 &\multicolumn{2}{c}{\mbox{''}}& 1.2 &\\
                  &\multicolumn{2}{c}{\mbox{''}}&\multicolumn{2}{c}{\mbox{''}}& 1411 &\multicolumn{2}{c}{\mbox{''}}& 1.9 &\\
\hline
Parkes       & 240 & 250 & -- 5 &  5                   & 2400 & 100 & 800 & 1.86 & Baccigalupi et al.\ 2001\\
Duncan et al.& 250 & 260 &\multicolumn{2}{c}{\mbox{''}}&  ''  &\multicolumn{2}{c}{\mbox{''}}& 1.86 & \\ 
(1997)       & 260 & 270 &\multicolumn{2}{c}{\mbox{''}}&  ''  &\multicolumn{2}{c}{\mbox{''}}& 1.43 & \\  
             & 270 & 280 &\multicolumn{2}{c}{\mbox{''}}&  ''  &\multicolumn{2}{c}{\mbox{''}}& 1.67 & \\   
             & 280 & 290 &\multicolumn{2}{c}{\mbox{''}}&  ''  &\multicolumn{2}{c}{\mbox{''}}& 1.91 & \\   
             & 290 & 300 &\multicolumn{2}{c}{\mbox{''}}&  ''  &\multicolumn{2}{c}{\mbox{''}}& 1.77 & \\   
             & 300 & 310 &\multicolumn{2}{c}{\mbox{''}}&  ''  &\multicolumn{2}{c}{\mbox{''}}& 1.28 & \\   
             & 310 & 320 &\multicolumn{2}{c}{\mbox{''}}&  ''  &\multicolumn{2}{c}{\mbox{''}}& 1.64 & \\   
             & 320 & 330 &\multicolumn{2}{c}{\mbox{''}}&  ''  &\multicolumn{2}{c}{\mbox{''}}& 1.93 & \\   
             & 330 & 340 &\multicolumn{2}{c}{\mbox{''}}&  ''  &\multicolumn{2}{c}{\mbox{''}}& 1.44 & \\   
             & 340 & 350 &\multicolumn{2}{c}{\mbox{''}}&  ''  &\multicolumn{2}{c}{\mbox{''}}& 1.47 & \\   
             & 350 & 360 &\multicolumn{2}{c}{\mbox{''}}&  ''  &\multicolumn{2}{c}{\mbox{''}}& 1.49 & \\   
             & 240 & 360 &\multicolumn{2}{c}{\mbox{''}}&  ''  &\multicolumn{2}{c}{\mbox{''}}& 1.7  & Bruscoli et al.\ 2002 \\
             &326.5&331.5& -- 1 &  4 &  ''  &\multicolumn{2}{c}{\mbox{''}}& 1.68 & Tucci et al.\ 2002 \\ 
             & 240 & 360 & -- 5 &  5 &  ''  &  40 & 250 & 2.4  &  Giardino et al.\ 2002\\   
\hline
Effelsberg   &  20 &  30 & -- 5 &  5                   & 2695 & 100 & 800 & 1.79 & Baccigalupi et al.\ 2001 \\ 
Duncan et al.&  30 &  40 &\multicolumn{2}{c}{\mbox{''}}&  ''  &\multicolumn{2}{c}{\mbox{''}}& 1.72 & \\ 
(1999)       &  40 &  50 &\multicolumn{2}{c}{\mbox{''}}&  ''  &\multicolumn{2}{c}{\mbox{''}}& 1.55 & \\ 
             &  50 &  60 &\multicolumn{2}{c}{\mbox{''}}&  ''  &\multicolumn{2}{c}{\mbox{''}}& 1.98 & \\ 
             &  55 &  65 &\multicolumn{2}{c}{\mbox{''}}&  ''  &\multicolumn{2}{c}{\mbox{''}}& 1.93 & \\ 
             &   5 &  75 &\multicolumn{2}{c}{\mbox{''}}&  ''  &\multicolumn{2}{c}{\mbox{''}}& 1.6  & Bruscoli et al.\ 2002 \\ 
\hline
Effelsberg       &  45 &  55 &    5 & 20 & 1400 & 100 & 800 & 1.5  & Bruscoli et al.\ 2002\\
Uyan\i ker et al.& 140 & 150 &    4 & 10 &  ''  &\multicolumn{2}{c}{\mbox{''}}& 2.5  & \\
(1999)           & 190 & 200 &    8 & 15 &  ''  &\multicolumn{2}{c}{\mbox{''}}& 2.3  & \\
\hline
\multicolumn{10}{l}{\mbox{}}
\end{tabular}
\caption{Studies of power spectra of diffuse polarized intensity from
different surveys. Given are the range in Galactic longitude $l$ and
latitude $b$, the frequency of observation $\nu$, the range in
multipole $\ell$ for which the power spectrum was computed, the
spectral index of $P$ $\alpha_P$ and the reference to the study of the
power spectra.}
\label{t9:lit}
\end{table*}
\pagebreak
\begin{table*}
\centering
\begin{tabular}{lr@{ - }lr@{ - }lrr@{ - }lrl} 
 survey & \multicolumn{2}{c}{$l\;(\dg)$} & \multicolumn{2}{c}{$b\;(\dg)$} & $\nu$
     (MHz) & \multicolumn{2}{c}{$\ell$} & $\alpha_P$ & reference\\
\hline\hline
ATCA           & 327 & 331     & -- 0.5  & 3.5               & 1400 & 600 & 6000 & 1.68 & Tucci et al.\ 2002\\
Gaensler et al.& 329   & 332   & 0   & 3                     &  '' &\multicolumn{2}{c}{\mbox{''}}& 1.66 & \\
(2001)         & 326   & 329   &\multicolumn{2}{c}{\mbox{''}}&  '' &\multicolumn{2}{c}{\mbox{''}}& 1.68 & \\
               & 330.1 & 331.1 & 0.5 & 1.5                   &  '' &\multicolumn{2}{c}{\mbox{''}}& 1.79 & \\
               & 329.2 & 330.2 & 0.6 & 1.6                   &  '' &\multicolumn{2}{c}{\mbox{''}}& 2.24 & \\
               & 327.2 & 328.2 &\multicolumn{2}{c}{\mbox{''}}&  '' &\multicolumn{2}{c}{\mbox{''}}& 1.93 & \\
               & 328   & 329   & 2.1 & 3.1                   &  '' &\multicolumn{2}{c}{\mbox{''}}& 2.58 & \\
               & 326.5 & 327.5 & 1.8 & 2.8                   &  '' &\multicolumn{2}{c}{\mbox{''}}& 1.88 & \\
               & 329.9 & 330.9 &\multicolumn{2}{c}{\mbox{''}}&  ''
     &\multicolumn{2}{c}{\mbox{''}}& 1.68 & \\
\hline
WSRT & 158 & 165 &   13 & 20 &  350 & 100 & 1000 & 2.26 & Haverkorn et al.\ 2003a \\
\hline
WSRT & 134 & 141 &    3 & 10 &  350 & 100 & 1000 & 2.20 & Haverkorn et al.\ 2003b \\
\hline
WSRT & 156   & 143   & 0.5  & 7.5                  &  327 & 100 & 1500
     & 1.67 & Schnitzeler et al.\ in prep \\
     & 162   & 169   &  8   & 15                   & ''  &\multicolumn{2}{c}{\mbox{''}}& 1.48 &\\ 
     & 155.5 & 161.5 &\multicolumn{2}{c}{\mbox{''}}& ''  &\multicolumn{2}{c}{\mbox{''}}& 1.84 &\\ 
     & 147.5 & 154.5 &\multicolumn{2}{c}{\mbox{''}}& ''  &\multicolumn{2}{c}{\mbox{''}}& 1.70 &\\ 
     & 140.5 & 147.5 &\multicolumn{2}{c}{\mbox{''}}& ''  &\multicolumn{2}{c}{\mbox{''}}& 1.63 &\\ 
     & 139   & 146   & 15 & 22                     & ''  &\multicolumn{2}{c}{\mbox{''}}& 1.24 &\\ 
     & 146.5 & 153.5 &\multicolumn{2}{c}{\mbox{''}}& ''  &\multicolumn{2}{c}{\mbox{''}}& 0.99 &\\ 
     & 153.5 & 160.5 &\multicolumn{2}{c}{\mbox{''}}& ''  &\multicolumn{2}{c}{\mbox{''}}& 1.67 &\\ 
     & 153.3 & 160.5 & 22 & 29                     & ''  &\multicolumn{2}{c}{\mbox{''}}& 0.73 &\\ 
     & 147   & 154   &\multicolumn{2}{c}{\mbox{''}}& ''  &\multicolumn{2}{c}{\mbox{''}}& 0.90 &\\ 
     & 140.5 & 147.5 &\multicolumn{2}{c}{\mbox{''}}& ''  &\multicolumn{2}{c}{\mbox{''}}& 0.68 &\\ 

\hline
\multicolumn{10}{l}{\mbox{}}
\end{tabular}

{\bf Table 3.} {\it Continued}
\end{table*}

\end{document}